\documentclass[11pt,onecolumn,draftcls]{./IEEEtran/IEEEtran}
\usepackage{amssymb}
\usepackage[cmex10]{amsmath}
\usepackage{amsthm}
\usepackage{algorithmic}
\usepackage{algorithm}
\usepackage{graphicx}

\newtheorem{thm}{Theorem}

\renewcommand{\Re}{\mathbb{R}}

\providecommand{\abs}[1]{\lvert#1\rvert}
\providecommand{\norm}[1]{\lVert#1\rVert}

\ifCLASSINFOpdf
\else
\fi

\begin{document}
\title{Multiscale Gossip for Efficient Decentralized Averaging in Wireless Packet Networks}

\author{Konstantinos I.~Tsianos and Michael G.~Rabbat% <-this % stops a space
\thanks{K.I.~Tsianos and M.G.~Rabbat are with the Department
of Electrical and Computer Engineering, McGill University, Montr\'eal,
Qu\'ebec, H3A 2A7 Canada, e-mail: konstantinos.tsianos@mail.mcgill.ca, michael.rabbat@mcgill.ca.}% <-this % stops a space
\thanks{Parts of this work were presented at the 2010 International Conference on Distributed Computing in Sensor Systems; see~\cite{tsianos10}.}%
}

\maketitle

\begin{abstract}
This paper describes and analyzes a hierarchical algorithm called Multiscale Gossip for solving the distributed average consensus problem in wireless sensor networks. The algorithm proceeds by recursively partitioning a given network into subnetworks. Initially, nodes at the finest scale gossip to compute local averages. Then, using multi-hop communication and geographic routing to enable gossip between nodes that are not directly connected, these local averages are progressively fused up the hierarchy until the global average is computed. We show that the proposed hierarchical scheme with $k=\Theta(\log \log n)$ levels of hierarchy is competitive with state-of-the-art randomized gossip algorithms in terms of message complexity, achieving $\epsilon$-accuracy with high probability after $O\big(n \log \log n \log \frac{1}{\epsilon} \big)$ messages. Key to our analysis is the way in which the network is recursively partitioned. We find that the optimal scaling law is achieved when subnetworks at scale $j$ contain $O(n^{(2/3)^j})$ nodes; then the message complexity at any individual scale is $O(n \log \frac{1}{\epsilon})$. Another important consequence of hierarchical construction is that the longest distance over which messages are exchanged is $O(n^{1/3})$ hops (at the highest scale), and most messages (at lower scales) travel shorter distances. In networks that use link-level acknowledgements, this results in less congestion and resource usage by reducing message retransmissions. Simulations illustrate that the proposed scheme is more message-efficient than existing state-of-the-art randomized gossip algorithms based on averaging along paths.
\end{abstract}

%\begin{IEEEkeywords}
%Distributed averaging, distributed signal processing, gossip algorithms, hierarchical decomposition, multi-tier architecture.
%\end{IEEEkeywords}

% \fi
\IEEEpeerreviewmaketitle

\section{Introduction}
\label{sec:intro}
Distributed signal and information processing applications arise in a variety of contexts including wireless sensor networks, the smart-grid, large-scale unmanned surveillance, and mobile social networks.  Large-scale applications demand protocols and algorithms that are robust, fault-tolerant, and scalable.  Energy-efficiency is also an increasingly important design factor.  When a system is comprised of battery-powered nodes or agents equipped with wireless radios for transmission---such as in wireless sensor networks---energy-efficiency equates to requiring few transmissions since in addition to consuming bandwidth, each wireless transmission dissipates battery resources.

Gossip algorithms~\cite{Boyd06,kempe03,Tsitsiklis86,Tsitsiklis84,dimakisSurvey} are an attractive paradigm for decentralized, in-network processing, and have received much attention in the computer science, systems and control, information theory, and signal processing research communities of late.  Gossip algorithms are frequently posed and studied as solutions to the \emph{distributed averaging} problem: in a network of $n$ nodes whose topology is described by a graph $G = (V, E)$ with $|V|=n$, each node initially has a scalar value $x_i(0)$, and the goal is to approximate the average, $x_{\mbox{\scriptsize ave}} = \frac{1}{n}\sum_{i=1}^n x_i(0)$ at every node.  Nodes iteratively and asynchronously exchange estimates with a small subset of the entire network, updating their local estimate after each exchange.  These protocols have a number of attractive properties.  The simplicity of the protocol (exchange information, update, repeat) makes it extremely robust; since there is no fixed routing of information to a fusion center and since all nodes compute a solution, there is no single point of failure or bottleneck.  Furthermore, past studies have demonstrated that gossip algorithms converge even under unreliable or dynamic networking conditions; see, e.g.,~\cite{dimakisSurvey} and references therein.

However, the standard gossip algorithms for distributed averaging~\cite{Boyd06,kempe03,Tsitsiklis86,Tsitsiklis84} constrain information to only be exchanged between neighboring nodes and exhibit poor scaling and energy-efficiency in topologies frequently used to model connectivity in wireless networks, such as grids and random geometric graphs~\cite{gupta98}.  Roughly speaking, the number of messages transmitted per node depends linearly on $n$, the size of the network.  Since the $n$ values required to compute the average are initially stored at different nodes, any distributed averaging algorithm requires that each node perform at least one transmission.  This discrepancy between constant and linear transmissions per node, has motivated the development of a number of variants of gossip algorithms specifically aimed at improving the efficiency of gossip on grid and random geometric graph topologies (see Section~\ref{sec:previous} for more).

The principle of hierarchical (multiscale) decomposition, or divide-and-conquer, arises in a variety of settings as a mechanism which yields efficient information processing procedures.  In the signal processing and coding communities, multiscale analysis is frequently associated with wavelet-based methods, e.g., for signal and image denoising, edge detection, and transform coding~\cite{mallat}.  A hierarchical approach to communication over wireless networks was shown to achieve the optimal capacity scaling law~\cite{ozgur07}.  A recent study also found that flocks of birds exhibit hierarchical organization and suggested that hierarchical behavior has been selected (in the evolutionary sense) because it is more efficient than democratic or individualistic strategies~\cite{nagy10}.

This paper describes and analyzes a multiscale gossip algorithm for distributed averaging in grid and random geometric graph topologies.  The network is recursively partitioned into smaller subnetworks.  The size of subnetworks at each scale and the number of scales of the partition depend on the number of nodes in the network.  Multiscale gossip operates over this partition in a bottom-up fashion.  First, all nodes within each subnetwork at the finest scale gossip until computing a suitably accurate local average.  Then, a representative node is elected for each subnetwork; an overlay grid is formed among all representatives within the same subnetwork at the next higher scale, and the representatives gossip over the overlay grid.  This procedure is repeated until representatives at the coarsest scale have computed an accurate approximation to the network average. At that point the representatives disseminate their estimate to all of their children in the hierarchy.  Multi-hop communication between representatives at coarser scales is accomplished using geographic routing \cite{Dimakis08,sarkar09greedy}.

Our main contribution is the analysis of multiscale gossip.  In particular, for a carefully designed multiscale partition, we show that the total number of single-hop transmissions required to reach a desired level of accuracy $1/n$ scales nearly-linearly, requiring $O(n \log \log n \log n)$ total transmissions as $n \rightarrow \infty$ on random geometric graph and grid topologies. Consequently, the average number of transmissions per node is $O(\log \log n \log n)$.  Since information dissemination (randomized broadcast) is much more efficient than gossip (which is a form of information diffusion) in these topologies, representatives at all scales can optionally disseminate intermediate results to other nodes in their subnetwork, thereby improving robustness and fault-tolerance of the scheme, without affecting the order-wise scaling law.  In contrast to \emph{geographic gossip with path averaging}~\cite{benezit07}, a randomized gossip scheme with a linear scaling that also uses geographic routing to exchange information over multiple hops, multiscale gossip requires fewer and shorter multi-hop transmissions; for example, in a $\sqrt{n}\times \sqrt{n}$ grid topology, path averaging requires relaying messages over $O(n^{1/2})$ hops, whereas multiscale gossip messages at the coarsest scale are relayed over at most $O(n^{1/3})$ hops, and messages at finer scales travel significantly shorter distances.  This has advantages when reliable transmission (i.e., handshaking, forward error-correcting, and/or retransmission) protocols are used at the link-level to ensure accurate reception over each link of a multi-hop path.  Moreover, at each iteration of multiscale gossip, information is only exchanged between one pair of nodes, as opposed to all nodes along a path.

The remainder of this paper is organized as follows.  Section~\ref{sec:previous} covers background, and related work. Section~\ref{sec:hgossip} describes the procedure for recursively constructing the hierarchical network partition and for carrying out multiscale gossip. Then, our main results are presented in Section~\ref{sec:results}, with analysis and proofs provided in Section~\ref{sec:analysis}. A numerical evaluation of the proposed algorithm is presented in Section~\ref{sec:evaluation}. Some practical considerations are discussed in Section~\ref{sec:consider}, and we conclude in Section~\ref{sec:future}.

\section{Background and Problem Definition}
\label{sec:previous}
Our primary measure of performance is \emph{communication cost}---the number of messages (single hop transmissions) required to compute an estimate to $\epsilon$ accuracy---which is also considered in~\cite{Dimakis08,benezit07}.  Moreover, we are interested in characterizing scaling laws, or the rate at which the communication cost increases as a function of network size.  In the analysis of scaling laws for gossip algorithms, a commonly studied measure of convergence rate is the $\epsilon$-averaging time, denoted $T_\epsilon(n)$ and defined as~\cite{Boyd06}
\begin{eqnarray}
T_{\epsilon}(n) = \sup_{x(0)} \inf \left\{t\ :\ \Pr\left(\frac{\|x(t) - x_{\mbox{\scriptsize ave}}\|}{\|x(0)\|} \ge \epsilon\right) \le \epsilon\right\},
\end{eqnarray}
which is the number of iterations required to reach an estimate with $\epsilon$ accuracy with high probability.  The $\epsilon$-averaging time $T_{\epsilon}(n)$ reflects the idea that the complexity of gossiping on a particular class of network topologies should depend both on the final accuracy and the network size.  When only neighbouring nodes communicate at each iteration, $T_\epsilon(n)$ and communication cost are identical up to a constant factor.  Otherwise, communication cost can generally be bounded by the product of $T_\epsilon(n)$ and a bound on the number of messages required per iteration.

In wireless sensor network applications, \emph{random geometric graphs} are a typical model for connectivity since communication is restricted to nearby nodes.  In the $2$-dimensional random geometric graph model, $n$ nodes are randomly assigned coordinates uniformly in the unit square, and two nodes are connected with an edge when their Euclidean distance is less than or equal to a connectivity radius, $r(n)$~\cite{gupta98,penroseRGG}.  In~\cite{gupta98} it is shown that if the connectivity radius scales as $r_{\mbox{\scriptsize con}}(n) = \Theta(\sqrt{\frac{\log n}{n}})$ then the network is connected with high probability.  Throughout this paper when we refer to a random geometric graph, we mean one with the connectivity $r_{\mbox{\scriptsize con}}(n)$.  

Although the standard neighbor gossip algorithms are known to be efficient on complete graphs and expander-like topologies, they are also known to converge slowly on grids and random geometric graphs, two topologies commonly used to model wireless networks~\cite{kempe03,Boyd06}.  Kempe, Dobra, and Gehrke~\cite{kempe03} initiated the study of scaling laws for gossip algorithms and showed that gossip requires $\Theta(n \log \epsilon^{-1})$ total messages to converge on complete graphs.  Boyd, Ghosh, Prabhakar, and Shah~\cite{Boyd06} studied scaling laws for standard randomized gossip on random geometric graphs and found that communication cost scales as $\Theta(\frac{n^2}{\log n} \log \epsilon^{-1})$ messages even if the algorithm is optimized with respect to the topology.  This finding motivated the pursuit of efficient gossip algorithms for wireless networks in a number of interesting directions.  For a complete overview of this line of work, we refer the reader to the recent survey~\cite{dimakisSurvey}.  Here we briefly discuss different approaches, focusing on advances most closely related to the present article.

A number of approaches seek more efficient computation while enforcing the constraint that information only be exchanged between neighboring nodes at each iteration. Most of these approaches introduce memory at each node, creating higher-order updates similar to shift-registers or polynomial filters~\cite{Cao06,Kok09}.  Scaling laws for a deterministic, synchronous variant of this approach are presented in~\cite{Oreshkin10}, leading to $\Theta(\frac{n^{1.5}}{\sqrt{\log n}} \log \epsilon^{-1})$ communication cost.  Related asynchronous gossip algorithms based on lifted Markov chains have been proposed that achieve similar scaling laws~\cite{LiDaiIT,jung07}.  Recent work~\cite{Jung09} suggests that no gossip algorithm on grids and random geometric graphs can achieve better than $O(n^{1.5} \log \epsilon^{-1})$ scaling while constraining information exchange to be solely between neighboring nodes.

A variant called \emph{geographic gossip}, proposed by Dimakis, Sarwate, and Wainwright~\cite{Dimakis08}, achieves a communication cost of $\Theta(\frac{n^{1.5}}{\sqrt{\log n}} \log \epsilon^{-1})$ by allowing distant (non-neighbouring) pairs of nodes to gossip at each iteration.  Assuming that each node knows its own coordinates and the coordinates of its neighbours in the unit square, communication between arbitrary pairs of nodes is made possible using \emph{greedy geographic routing}.  Rather than addressing nodes directly, a message is sent to a randomly chosen target $(x,y)$-location, and the recipient of the message is the node closest to that target.  To reach the target, a message is forwarded from a node to its neighbour who is closest to the target. If a node is closer to the target than all of its neighbours, this is the final message recipient.  It is shown in~\cite{Dimakis08} that for random geometric graphs with connectivity radius $r(n) = r_{\mbox{\scriptsize con}}(n)$, greedy geographic routing succeeds with high probability.  For an alternative form of greedy geographic routing, which may be useful in implementations see~\cite{sarkar09greedy}.  The main contribution of~\cite{Dimakis08} is to illustrate that allowing nodes to gossip over multiple hops can lead to significant improvements in message cost.  In follow-up work, Benezit, Dimakis, Thiran, and Vetterli~\cite{benezit07} showed that a modified version of geographic gossip, called \emph{path averaging}, can achieve $\Theta(n \log \epsilon^{-1})$ message cost on random geometric graphs.  To do this, all nodes along the path from the source to the target participate in a gossip iteration. If geographic routing finds a path through nodes $S = \{x_{i}, \ldots, x_{j}\}$ to deliver a message from $x_{i}$ to $x_{j}$, the estimates of all nodes in $S$ are accumulated on the way to $x_{j}$. Then $x_{j}$ computes the average of all $|S|$ values and sends the average back down the same path towards $x_{i}$, and all nodes in $S$ update their estimates. 

Observe that there is a tradeoff between algorithmic simplicity and performance. If we only allow pairwise communication between neighboring nodes, we cannot beat the $O(n^{1.5} \log \epsilon^{-1})$ barrier. On the other hand, if we have the additional knowledge of geographical information for each node and its immediate neighbours, we can use geographic routing and with the added complexity of averaging over paths we can bring the message complexity down to linear at the expense of messages having to travel potentially over $O(n^{1/2})$ hops. However, in order to improve upon the performance achievable using pairwise communication between neighboring nodes, some additional complexity must be introduced. In this work, rather than averaging along paths, we propose to decompose computation in a multiscale manner in order to achieve faster convergence.

The multiscale approach considered in this paper also assumes that the nodes know their own and their neighbour's coordinates in the unit square. Using the geographic information, we derive a hierarchical algorithm that asymptotically achieves a communication cost of $O(n \log \log n \log \epsilon^{-1})$ messages, which is equivalent to that of path averaging up to a logarithmic factor. However, in multiscale gossip, information is only exchanged between pairs of nodes, and there is no averaging along paths. At the expense of extra complexity for building the logical hierarchy, besides near-optimal communication cost we achieve two other important goals. First, the longest distance a message travels in our multiscale approach is $O(n^{1/3})$ hops which is much shorter compared to $O(n^{1/2})$ hops for geographic gossip or path averaging. This can prove significant if an adversary wishes to disrupt gossip computation by forcing the network to drop a particular message or by deactivating a node in the middle of an iteration. In that scanario a substantial amount of information can be lost in path averaging since each iteration involves $O(\sqrt{\frac{n}{\log n}})$ nodes on average. Second, as we show later on, multiscale gossip distributes the computation quite evenly across the network and does not overwhelm and deplete the nodes located closer to the center of the unit square as is the case for path averaging.

We note that we are not the first to propose gossiping in a multiscale or hierarchical manner.  Sarkar et al.~\cite{gossip07sarkar} describe a hierarchical approach for computing aggregates, including the average.  However, because their algorithm uses order and duplicate insensitive synopses to estimate the desired aggregate, the size of each message exchanged between a pair of nodes must scale with the size of the network.  Other hierarchical distributed averaging schemes that have been proposed in the literature focus on the synchronous form of gossip, and they do not prove scaling laws for communication cost, nor do they provide rules for forming the hierarchy (i.e.  assume the hierarchical decomposition is given) \cite{kim08,epstein08,cattivelli09}. Finally, we mention that hierarchical approaches to routing have also been proposed \cite{DoublingMetrics,HierRoutingDynNets}. Although this line of work uses similar techniques (hierarchy and divide-and-conquer approach) the problems considered are not related to distributed averaging.

A preliminary version of this work appears in the conference paper~\cite{tsianos10} where multiscale gossip is described, including our construction of the multiscale network partition and a simple communication cost analysis.  The present manuscript extends~\cite{tsianos10} in several ways. In addition to more detailed communication cost analysis, the proof of the final error bound of multiscale gossip is a new result. Moreover, we investigate the role of the subdivision parameter $a$, explaining in what sense the value $a=\frac{2}{3}$ is optimal. Finally, we include a thorough set of experiments to evaluate the performance of multiscale gossip. 

% \section{Network Model and Problem Definition}
% \label{sec:model}
% \input{model}

\section{Multiscale Gossip}
\label{sec:hgossip}
Multiscale gossip performs averaging in a hierarchical manner. At each moment only nodes in the same level of hierarchy do computations at a local scale and  computation at one level begins after the previous level has finished. By hierarchically decomposing the initial graph into subgraphs, we impose an order in the computation. As shown in the next section, for a specific decomposition it is possible to divide the overall work into a small number of linear sub-problems and thus  obtain very close to linear complexity in the size of the network.

\begin{figure} 
\begin{center}
\includegraphics[height=5cm]{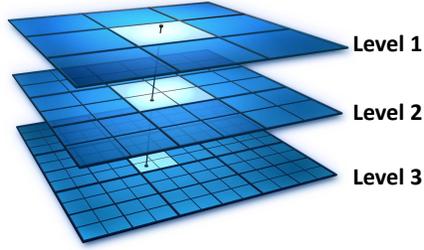}
\caption{\label{fig:multigrid} Hierarchical multiscale subdivision of the unit square. At each level, each cell is split into equal numbers of smaller cells. Before the representatives of the cells can gossip on a grid graph, we run gossip on each cell. } 
\end{center}
\end{figure}

Assume we have a random geometric graph $G=(V,E)$ and each node knows its own coordinates in the unit square and the locations of its immediate neighbours. Each node also knows the total number of nodes $n$ in the network and $k$, the desired number of hierarchy levels\footnote{As explained in Section \ref{sec:analysis}, given $n$, the number of levels $k$ can be computed automatically.}. Figure \ref{fig:multigrid} illustrates an example with $k=3$. We use the convention that level $k$ is the lowest level where the unit square is split into many small cells. Level $1$ is the top level where we only have few big cells. All cells at the same level have the same area. The way we split each cell into subcells is directed by a subdivision constant $a=\frac{2}{3}$ whose value is justified in Section \ref{sec:analysis}. If a cell contains $n$ nodes, it is split into $n^{1-a}$ cells of dimensions $n^\frac{a-1}{2} \times n^\frac{a-1}{2}$ each. At level $k$ the unit square is split into $d_{k}$ cells $C_{(k,1)},\ldots, C_{(k,d_k)}$. Each cell $C_{(k,\cdot)}$ contains nodes forming a subgraph of $G$. Each subgraph $G_{(k,\cdot)}$ runs standard randomized gossip until convergence. Then, in each subgraph $G_{(k,\cdot)}$ we elect one representative node $L_{(k,\cdot)}$. The representative selection can be randomized or deterministic as explained in Section \ref{sec:consider}. Generally, not all subgraphs have the same number of nodes. For this reason, the value of each representative has to be reweighted proportionally to its subgraph's size. At level $k-1$, the unit square is split into $d_{k-1}$ cells $C_{(k-1,1)},\ldots, C_{(k-1,d_{k-1})}$. Each cell $C_{(k-1,\cdot)}$ contains the same number of $C_{(k,\cdot)}$ cells. The representatives of the $G_{(k,\cdot)}$ graphs are then organized into logical grid graphs $G_{(k-1,1)},\ldots,G_{(k-1,d_{k-1})}$, one grid per $C_{(k-1,\cdot)}$ cell. Two representatives $L_{(k,i)}$ and $L_{(k,j)}$ are connected by an edge in a $G_{(k-1,\cdot)}$ if cells $C_{(k,i)}$ and $C_{(k,j)}$ are adjacent and contained in the same $C_{(k-1,\cdot)}$ cell. Note that representatives can determine which cells they are adjacent to given the current level of hierarchy and $n$ since the cell construction is deterministic. Next, we run randomized gossip simultaneously on all $G_{(k-1,\cdot)}$ grid graphs. Finally, we select a representative node $L_{(k-1,\cdot)}$ for each $G_{(k-1,\cdot)}$ grid graph and continue the next  hierarchy level. The process is repeated until we reach level $1$ at which point we have only one grid graph $G_{(1,1)}$ contained in the single cell $C_{(1,1)}$. By construction $C_{(1,1)}$ coincides with the unit square. Once randomized gossip on $G_{(1,1)}$ is over, each node of $G_{(1,1)}$ as a representative $L_{(2,\cdot)}$ disseminates its final value to all the nodes in its $C_{(2,\cdot)} $ cell.

% \footnote{Note that at level $k$ as well as any other level $d$ there are many $C_{d}$ cells but for simplicity we avoid denoting them  $C_{d,i}$ with $i$ running from $1$ to maximum number of cells. Similarly for representatives $L_{d}$. } . 

\begin{algorithm} 
\caption{MultiscaleGossip($x_{init}, C, n, q, k, \epsilon$)} 
\label{alg:hiergossip}
\begin{algorithmic}[1] 
\STATE $a = \frac{2}{3}$
\IF{$q < k$}
\STATE Split $C$ into $m_{q+1} = n^{1-a}$ cells: $C_{(q+1,1)},\ldots,C_{(q+1,m_{q+1})}$
\STATE Select a representative node $L_{(q+1,i)}$ for each cell $C_{(q+1,i)}, i \in \{1,\ldots,m_{q+1}\}$
\FORALL{cells $C_{(q+1,i)}$}
\STATE \textbf{call} \textit{MultiscaleGossip($x_{init}(C_{(q+1,i)}), C_{(q+1,i)}, n^{a}, q+1, tol$)}
\ENDFOR
\STATE Form grid graph $G_{(q,\cdot)}$ of representatives  $L_{(q+1,i)}$
\STATE \textbf{call} \textit{RandomizedGossip($x_{init}(L_{(q+1,1:m_{q+1})}), G_{(q,\cdot)}, \epsilon$)}
\IF{q = 1}
\STATE Spread value of $L_{(2,i)}$ to all nodes in each $C_{(2,i)}$
\ENDIF
\ELSE
\STATE Form graph $G_{(k,\cdot)}$ only of nodes in $V(G)$ contained in $C$
\STATE \textbf{call} \textit{RandomizedGossip($x_{init}, G_{(k,\cdot)}, \epsilon$)}
\STATE Reweight representative values as : $x(L_{(k,i)}) = x(L_{(k,i)}) \frac{|V(G_{k})| \cdot m_{k-1} }{|V(G)|}$
\ENDIF
\end{algorithmic}
\end{algorithm}

Algorithm \ref{alg:hiergossip} describes multi-scale gossip in a recursive manner. The initial call to the algorithm has as arguments, the vector of initial node values ($x_{init}$), the unit square ($C=[0,1] \times [0,1]$), the network size $n$, the top level $q=1$, the desired number of hierarchy levels $k$ and the desired error tolerance $\epsilon$ to be used by each invocation of randomized gossip. In a down-pass the unit square is split into smaller and smaller cells all the way to the $C_{(k,\cdot)}$ cells. After gossiping in the $G_{(k,\cdot)}$ subgraphs in \textit{Line 15}, the representatives adjust their values (\textit{Line 16}). As explained in the next section, if $k$ is large enough, each $G_{(k,\cdot)}$  is a complete graph. Since each node knows the locations of its immediate neighbours (needed for geographic routing), at level $k$ we can also compute the size of each $G_{(k,\cdot)}$ graph which is needed for the reweighting. The up-pass begins with the $L_{(k,\cdot)}$ representatives forming the $G_{(k-1,\cdot)}$ grid graphs (\textit{Line 8}) and then running gossip in all of them in parallel. Between consecutive levels we use a parameter $a=\frac{2}{3}$ to decide how many $C_{(r+1,\cdot)}$ cells fit in each $C_{(r,\cdot)}$ cell. As mentioned earlier, the motivation for this parameter and its specific value is explained in the sequel. Notice the pseudocode mimics a sequential single processor execution which is in line with the analysis that follows in Section \ref{sec:analysis}. However, it should be emphasized that the algorithm is intended for and can be implemented in a distributed fashion. The notation $x_{init}(C)$ or $x_{init}(L)$ indicates that we only select the entries of $x_{init}$ corresponding to nodes in cell $C$ or representatives $L$.

\section{Main Results}
\label{sec:results}
Before proceeding with the detailed analysis we present here our main results. Proofs are provided in Section~\ref{sec:analysis} below. Section \ref{sec:hgossip} above describes multiscale gossip, an algorithm for distributed average consensus on random geometric graphs which uses randomized gossip as a black box. If each invocation of randomized gossip runs up to $\epsilon$ accuracy, the total number of messages used by multiscale gossip is given in Theorem~\ref{th:msgcomplexity}.

\begin{thm} \label{th:msgcomplexity}
Let a random geometric graph $G$ of size $n$ and constant $\epsilon > 0$ be given.  As the graph size $n \rightarrow \infty$, the communication cost of the multiscale gossip scheme described above with scaling constant $\alpha = \frac{2}{3}$ behaves as follows:
\begin{enumerate}
\item If the number of hierarchy levels $k$ remains fixed as $n \rightarrow \infty$, then the communication cost of multiscale gossip is $O\big(((k-1) n + n^{1 + (\frac{2}{3})^{k-1}}) \log \epsilon^{-1}\big)$ messages.
\item If the number of levels grows according to $k = \Theta(\log \log n)$ as $n \rightarrow \infty$, then the communication cost of multiscale gossip is $O(n \log \log n \log \epsilon^{-1})$ messages.
\end{enumerate}
\end{thm}

Note that $\epsilon$ in the theorem above is the target level of relative accuracy used each time we run randomized gossip on one overlay network, and not to the level of accuracy of the final average. Errors at intermediate levels can accumulate, but this accumulation is not catastrophic. An upper bound on this final accuracy is given in Theorem~\ref{th:msgerror} below.

\begin{thm} \label{th:msgerror}
Let a random geometric graph $G$ with $n$ nodes and initial values on the nodes $\overline{x}$ be given. If we run multiscale gossip on $G$ using $k$ levels of hierarchy and demand $\epsilon$-accuracy for randomized gossip at each subgraph, then with probability at least $(1-\epsilon)^g$, where $g$ is the total number of invocations of randomized gossip, the final error is not more than $n\epsilon$ i.e., 
\begin{equation}
 \frac{\norm{\overline{x}_{final} - \overline{x}_{av} }}{\norm{\overline{x}}} \leq n \epsilon,
\end{equation}
where $\overline{x}_{final} \in \Re^V$ denotes the vector of final estimates at each node, and $\overline{x}_{av}$ denotes a $n$-vector whose entries are all set to the average of the initial values at each node.
\end{thm}

The bound in Theorem~\ref{th:msgerror} is loose in two senses as explained in the end of its proof; namely, both the error bound is loose, and the probability with which the result holds is loose. The recursive partitioning scheme produces $g = 1 + n^{1-a} + n^{1-a^{2}} + ...+ n^{1 - a^{k-1}} = O(kn)$ graphs and one invocation of randomized gossip for each. We can control both the accuracy and the probability of success by carefully setting the value of $\epsilon$, the accuracy used each time we invoke randomized gossip within the overall multiscale gossiping procedure. If we want the final accuracy of multiscale gossip to be $\delta$ with high probability, we set the required accuracy for each randomized gossip call to $\epsilon = \frac{\delta}{k n}$. This will yield final accuracy $\frac{\delta}{k}$ which is in fact better than required. Moreover, the probability of achieving this accuracy will be at least $(1 - \frac{\delta}{kn})^g \geq(1 - \frac{\delta}{kn})^{kn} \rightarrow e^{-\delta} \ge 1-\delta$ as $n \rightarrow \infty$. The adjustment in $\epsilon$ also affects the total number of transmissions, as per Theorem~\ref{th:msgcomplexity}. Specifically, multiscale gossip requires $O(n \log{\log{n}} \log{\frac{1}{\epsilon}}) = O(n \log{\log{n}} \log{\frac{kn}{\delta}}) = O(n \log{\log{n}} \log(k n) + n \log{\log{n}} \log{\frac{1}{\delta}})$. As we see the transmissions are only increased by a logarithmic factor. In particular, letting the number of levels of hierarchy scale as $k = \Theta(\log \log n)$ and taking $\delta = \frac{1}{n}$ yields an overall message complexity of $O\big(n \log\log n \log n\big)$.

Besides the above main theoretical results, we have compared multiscale gossip to path averaging which is a recent state-of-the-art linear complexity algorithm. The experiments presented in Section~\ref{sec:evaluation} suggest that multiscale gossip has superior performance for graphs of up to many thousands or nodes. We also include an evaluation in scenarios with unreliable transmissions.

\section{Analysis}
\label{sec:analysis}
\subsection{Proof of Theorem \ref{th:msgcomplexity}}

Suppose we run multiscale gossip (Algorithm \ref{alg:hiergossip}) on a random geometric graph $G=(V,E)$ with $|V|=n$ and transmission radius $r(n) = \sqrt{\frac{c \log{n}}{n}}$. The topmost cell in the partition hierarchy is the unit square which we call cell $C_{(1,1)}$. We partition $C_{(1,1)}$ down to $k$ levels. At the highest level (level 1), we split the unit square into $m_{2}=n^{1-a}$ cells $C_{(2,\cdot)}$ each of area $\frac{1}{n^{1-a}} = n^{a-1}$ and dimensions $n^{\frac{a-1}{2}} \times n^{\frac{a-1}{2}}$. Below we exaplain why $a=\frac{2}{3}$. In each cell $C_{(2,\cdot)}$ we select a representative node $L_{(2, \cdot)}$. Representative nodes at this level form an overlay grid where logical edges exist between representatives of adjacent cells. Messages over logical edges may need multi-hop transmissions since the representatives will generally be out of each others range. The partition process repeats recursively within each $C_{(2,\cdot)}$ cell and so on until we reach the bottom level $k$. 

In general, on a 2-D grid of $p$ nodes, randomized gossip requires $O(p^{2} \log \epsilon^{-1})$ messages to achieve accuracy $\epsilon$ with probability $1-\epsilon$ (e.g. see \cite{Boyd06}). The grid graph $G_{(1,1)}$ formed by the representatives of the $C_{(2,\cdot)}$ cells has $n^{1-a}$ nodes. By using an appropriately large constant $c$ in the transmission radius (e.g. $c=3$), the random geometric graph $G$ is \textit{geo-dense}~\cite{cover_mixing_ran_geom_ercal_07} which means that a patch of area $n^{a-1}$ contains $\Theta(n^{a-1} \cdot n) = \Theta(n^{a})$ nodes with high probability. The maximum distance between two nodes of $G_{(1,1)}$ is $\sqrt{5} n^{\frac{a-1}{2}} = O(n^{\frac{a-1}{2}})$. To see this compute the maximum possible distance between two nodes in adjacent $C_{(2,\cdot)}$ cells using the Pythagorean theorem. If we divide by $r(n)$, we get a worst case estimate of the cost for multi-hop messages between representatives at level $1$: 
\begin{align}
MsgCost_{1} = O\left(\frac{ n^{\frac{a-1}{2}}}{\sqrt{\frac{c \log{n}}{n}}} \right) = O(n^{\frac{a}{2}})\ \text{single hop transmissions}. 
\end{align}
Notice that we have ignored the factor of $\sqrt{c \log{n}}$ thus slighlty overestimating the message cost. We do this to simplify the analysis and get a clean expression which allows us to compute the subdivision constant $a$. Knowing the cost of one (mutli-hop) message at level $1$ and the size of the grid $G_{(1,1)}$, the total number of single-hop transmissions for randomized gossip to converge on $G_{(1,1)}$ will be $O((n^{1-a})^{2} \log{\epsilon^{-1}}) \cdot O( n^{\frac{a}{2}}) = O(n^{2 - \frac{3a}{2}} \log{\epsilon^{-1}})$ which is $O(n \log{\epsilon^{-1}})$ if $a=\frac{2}{3}$.

Next let us look at the cost at the next level where we subdivide the cells $C_{(2,\cdot)}$. This will be instructive of how the process goes at any other level but the last. Each cell $C_{(2,\cdot)}$ contains $q=\Theta(n^{a})$ nodes and is subdivided into $q^{1-a'}$ cells $C_{(3,\cdot)}$  containing $q^{a'}$ nodes each (in expectation). Now, using again the geodensity property, a cell $C_{(3,\cdot)}$ containing $q^{a'}$ nodes will have area $\frac{q^{a'}}{n}$ and dimensions $\sqrt{\frac{q^{a'}}{n}} \times \sqrt{\frac{q^{a'}}{n}}$. Each cell $C_{(3,\cdot)}$ also contains a representative node $L_{(3,\cdot)}$. Following the same logic as before we can compute the cost of a message between the $L_{(3,\cdot)}$ representatives based on the worst possible single hop distance of nodes contained in adjacent cells $C_{(3,\cdot)}$. We just divide the cell dimension by $r(n)$ omitting the logarithmic factor to get $MsgCost_{2} = O(q^{\frac{a'}{2}})$. 

We have $n^{1-a}$ grid graphs $G_{(2,\cdot)}$. With $a=\frac{2}{3}$ and $q = \Theta(n^a)$, to make the total number of transmissions at level $2$ linear, we must have $n^{1-a} \cdot O((q^{1-a'})^{2} \log{\epsilon^{-1}}) \cdot O(q^{\frac{a'}{2}}) = O(n \log{\epsilon^{-1}}) \Rightarrow a' = \frac{2}{3}$ as well. In general, at any intermediate level $j$ the total number of transmissions is the number of grids, times the number of messages per grid, times the number of single hop transmissions per message to get between neighbouring representative nodes. Based on the above logic and using the same subdivision parameter $a$ at all levels, the expression for the cost at some level $j$ is 
\begin{equation}
 n^{1-a^{j-1}} \cdot O\left( \left( \left( n^{a^{j-1}} \right)^{1-a}\right)^2 \log \frac{1}{\epsilon} \right) \cdot O\left(\left( n^{a^{j-1}}\right)^{\frac{a}{2}} \right)
\end{equation}
which is linear in $n$ if $a=\frac{2}{3}$. Finally, we need to treat the last level which is $k$. At the last level we no longer have grids formed by representatives. Instead, the algorithm runs randomized gossip on each subgraph of $G$ with nodes contained inside each of the $C_{(k,\cdot)}$ cells. We have $O(n^{1 - (\frac{2}{3})^{k-1}})$ cells $C_{(k,\cdot)}$, each containing $n^{(\frac{2}{3})^{k-1}}$ nodes which are close enough to communicate via single hop messages. Since we run randomized gossip on each subgraph, the total number of messages at the last level is $O(n^{1+(\frac{2}{3})^{k-1}})$. Summing up all levels, plus $n$ messages to spread the final result back to all nodes, the total number of messages for multiscale gossip is $O\big( ((k-1) n +  n^{1+(\frac{2}{3})^{k-1}}) \log \epsilon^{-1}) + n\big) = O\big( ((k-1) n +  n^{1+(\frac{2}{3})^{k-1}}) \log \epsilon^{-1})\big)$. 

For the second part of the theorem, observe that at level $k$ each cell contains a subgraph of  $n^{(\frac{2}{3})^{k-1}}$ nodes in expectation. For constants $m \ge 2$ and $M \ge m$, we can choose $k$ so that each cell at the finest scale contains between $m$ and $M$ nodes with high probability, so that the cost per cell is bounded by $M^{2} \log \epsilon^{-1}$. In other words, choose $k$ such that $m \leq n^{(\frac{2}{3})^{k-1}} \leq M$, implying that $k = \Theta(\log\log n)$. Since the cost per cell at level $k$ is now bounded by a constant for $k = \Theta(\log \log n)$, the total level $k$ cost is $O(n^{1-(\frac{2}{3})^{k-1}} \log \epsilon^{-1})$ and the overall cost is $O\big( ((k-1)n +  n^{1-(\frac{2}{3})^{k-1}}) \log \epsilon^{-1}) + n\big) = O(n \log \log n \log \epsilon^{-1})$, completing the proof.

\subsection{Proof of Theorem \ref{th:msgerror}}
To simplify the discussion, we assume that each subgraph at each level has the same number of nodes. By geo-density of the random geometric graph~\cite{cover_mixing_ran_geom_ercal_07}, the number of nodes in each cell concentrates quickly as $n$ grows. The same proof technique can be extended to the general case, at the expense of much more cumbersome notation.

If each subgraph at level $i$ has $L_i$ nodes then we have $n = \prod_{j=1}^k L_j$. We introduce some notation to analyze the procedure or error propagation in its general form. The initial vector of values is $(x_1, \ldots, x_n)$. It is convenient to rewrite each element in $x$ as   $x_{l_1 l_{2} \ldots  l_k}$ where $1 \leq l_i \leq L_i$. We will write the whole vector as  $\{x_{l_1 l_{2} \ldots  l_k}\}$ using brackets. We overload our notation to describe the values of nodes that gossip at any level. For example at level $j$ we have the node values $x_{l_1 l_{2} \ldots  l_j}$ where the number of indices is indicative of the level. We will use $*$ notation to signify the converged values after gossiping at any level. E.g., after gossiping at level $j$, the node values are transformed to $x^{*}_{l_1 l_{2} \ldots l_j }$. Moreover, to advance from level $j+1$ to level $j$ we need to select one node at each subgraph at level $j+1$ as a representative and promote its value to the next level. This means that $x_{l_1 l_{2} \ldots l_{j} } = x^{*}_{l_1 l_{2} \ldots l_{j} c }$ for some $c$ in the range $1 \leq c \leq L_{j+1}$. 

Let us also write $m_{l_1 l_{2} l_{j-1}}^{L_j}$ to denote the mean of the values in a subgraph at level $j$ i.e.,
\begin{equation} \label{eq:mean}
 m_{l_1 l_{2} \ldots l_{j-1}}^{L_j} = \frac{\sum_{s_{j}=1}^{L_j} x_{l_1 l_{2}\ldots l_{j-1} s_{j}}}{L_j}.
\end{equation}

To begin the proof we first  state the error bounds for each intermediate subgraph after running randomized gossip to $\epsilon$-accuracy. At level $j$ for each subgraph we have
\begin{equation} \label{eq:levelacc}
\frac{\norm{\{x^{*}_{l_1 l_{2} \ldots l_{j-1} 1:L_j }\} - \overline{m}_{l_1 l_{2} \ldots l_{j-1}}^{L_j}}}{\norm{\{x_{l_1 l_{2} \ldots l_{j-1} 1:L_j } \}}} \leq \epsilon.
\end{equation}
where $\overline{m}$ signifies a vector with all its element equal to $m$. Using the definition of the $2$-norm, the inequality should hold for each summand, and so,
\begin{eqnarray}
\epsilon \norm{\{x_{l_1 l_{2} \ldots l_{j-1} 1:L_j } \}} &\ge& \abs{x^{*}_{l_1 l_{2} \ldots l_{j-1} c } - m_{l_1 l_{2} \ldots l_{j-1}}^{L_j}} \\
&=& \abs{x_{l_1 l_{2} \ldots l_{j-1}} - m_{l_1 l_{2} \ldots l_{j-1}}^{L_j}},
\end{eqnarray}
where the last equality follows since $x^{*}_{l_1 l_{2} \ldots l_{j-1} c } = x_{l_1 l_{2} \ldots l_{j-1}}$. At the top level $1$ we have:
\begin{equation} 
\frac{\norm{\{ x_{1:L_1}^*\} - \overline{m}^{L_1}}}{\norm{\{ x_{1:L_1} \}}} \leq \epsilon. 
\end{equation}

We can obtain a bound by squaring both sides and using the definition of the norm, and a second bound is again obtained by observing that each summand must be less than the right hand side:
\begin{subequations}\label{eq:toplevelacc}
\begin{align}
\sum_{s_{1}=1}^{L_1} (x_{s_{1}}^* - m^{L_1})^2 & \leq \epsilon^2 \norm{\{ x_{1:L_1} \}}^2 \label{first} \\
\abs{x_{s_{1}}^* - m^{L_1}}  & \leq \epsilon \norm{\{ x_{1:L_1} \}}. \label{second}
\end{align}
\end{subequations}

Once level $1$ is finished, the final values are distributed to all the nodes that each node represents. We are interested to bound the following :
\begin{equation}
 \frac{ \norm{(x_1^*,\ldots,x_1^*, \ldots, x_{L_1}^*, \ldots, x_{L_1}^*)- \overline{x}_{av}}}{\norm{x}}
\end{equation}
where in the above expression each value $x_{l_{1}}^*$ is repeated $L_{k} \times \cdots \times L_2$ times. 

We start by bounding the squared expression for simplicity. Using the definition of the $2$-norm, adding and subtracting the mean at level $1$, expanding the quadratic term, and using the bound~\eqref{first},
\begin{eqnarray} \label{eq:mainbound}
\lefteqn{\frac{ \norm{(x_1^*,\ldots,x_1^*, \ldots, x_{L_1}^*, \ldots, x_{L_1}^*)- \overline{x}_{av}}^2}{\norm{x}^2}}\\
&=& \frac{ L_{k}  \dots  L_2 \sum_{s_{1}=1}^{L_1} (x_{s_{1}}^* - x_{av})^2}{\norm{x}^2}  \\
&=& \frac{ L_{k}  \dots  L_2 \sum_{s_{1}=1}^{L_1} (x_{s_{1}}^* - m^{L_1} + m^{L_1} - x_{av})^2}{\norm{x}^2}  \\
&=& \frac{ L_{k}  \dots  L_2 \left[ \sum_{s_{1}=1}^{L_1} (x_{s_{1}}^* - m^{L_1})^2  + 2 (m^{L_1} - x_{av}) \sum_{s_{1}=1}^{L_1}(x_{s_{1}}^* - m^{L_1}) + L_1 (m^{L_1} - x_{av})^2 \right] }{\norm{x}^2} \\
& \leq& \frac{ L_{k}  \dots  L_2 \left[ \epsilon^2 \norm{\{ x_{1:L_1} \}}^2  +  L_1 (m^{L_1} - x_{av})^2 \right] }{\norm{x}^2}.
\end{eqnarray}
We arrived at the last inequality after noticing that $\sum_{s_{1}=1}^{L_1}(x_{s_{1}}^* - m^{L_1}) = 0$. The reason is that randomized gossip does not change the average (and thus the sum) of the values at any time. So both for the initial and the converged values at level $1$ we have $\sum_{s_{1}=1}^{L_1}x_{s_{1}}^* = \sum_{s_{1}=1}^{L_1} x_{s_{1}}$. But due to equation \ref{eq:mean}, $\sum_{s_{1}=1}^{L_1}x_{s_{1}} = L_1 m^{L_1} = \sum_{s_{1}=1}^{L_1} m^{L_1}$ and so $\sum_{s_{1}=1}^{L_1} x_{s_{1}}^* = \sum_{s_{1}=1}^{L_1} m^{L_1}$. 

Next we focus on bounding the two parts of the above numerator separately. For details of the derivation please see the appendix. The final results are:
\begin{eqnarray}
\frac{\epsilon^2 \norm{\{ x_{1:L_1} \}}^2}{\norm{x}^2} &\leq& \epsilon^{2} \\ \label{eq:bound1}
%\frac{m^{L_1} - x_{av}}{\norm{x}}  &\leq& (k-1) \epsilon \\
%\frac{\sum_{s_{1}=1}^{L_1}(x_{s_{1}}^* - m^{L_1}) }{\norm{x}} &\leq& L_{1} \epsilon^{2} + \sqrt{\frac{L_{1}}{L_{2}}} L_{1} \epsilon \\
\frac{L_1 (m^{L_1} - x_{av})^2}{\norm{x}^2} &\leq& L_{1} (k-1)^{2} \epsilon^{2}.\label{eq:bound2}
\end{eqnarray}

Using \eqref{eq:bound1} and \eqref{eq:bound2} with \eqref{first},\eqref{second} we obtain the overall bound:
%\begin{equation}
%\begin{split}
%\begin{align}
\begin{eqnarray}
\lefteqn{\frac{ \norm{(x_1^*,\ldots,x_1^*, \ldots, x_{L_1}^*, \ldots, x_{L_1}^*)- \overline{x}_{av}}^2}{\norm{x}^2}}  \\
& \leq &\frac{ L_{k}  \dots  L_2 \left( \epsilon^2 \norm{\{ x_{1:L_1} \}}^2  +  L_1 (m^{L_1} - x_{av})^2 \right) }{\norm{x}^2}  \\
& \leq &L_{k}  \dots  L_2 \left(  \epsilon^{2} +   L_{1}(k-1)^{2} \epsilon^{2} \right)  \\
& \leq &L_{k}  \dots  L_2  \epsilon^{2} + n (k-1)^{2} \epsilon^{2}.
\end{eqnarray}

Finally, by bounding $L_{k}  \dots  L_2 < n$ and $(k-1)^2 < n$ we get
\begin{eqnarray}
\lefteqn{\frac{ \norm{(x_1^*,\ldots,x_1^*, \ldots, x_{L_1}^*, \ldots, x_{L_1}^*)- \overline{x}_{av}}^2}{\norm{x}^2} }\\
&\leq& n \epsilon^{2} + n^{2} \epsilon^{2} \leq  2 n^{2} \epsilon^{2} ,
\end{eqnarray}
and so we arrive at the bound
\begin{align}
 \frac{ \norm{(x_1^*,\ldots,x_1^*, \ldots, x_{L_1}^*, \ldots, x_{L_1}^*)- \overline{x}_{av}}}{\norm{x}} \leq \sqrt{2} n \epsilon. \label{finalerror}
\end{align}
%\end{split}
%\end{equation}

This bound will hold whenever all randomized gossip operations at intermediate subgraphs achieve $\epsilon$ accuracy. Any invocation of randomized gossip achieves $\epsilon$ accuracy with probability at least $1-\epsilon$ and all randomized gossip operations are independent of each other. If we have $g$ subgraphs total appearing during a run of multiscale gossip, the probability that we achieve final error $\sqrt{2} n \epsilon$ is at least $(1 - \epsilon)^{g}$.

Notice that this bound is relatively loose. This should be expected given it was obtained using very loose bounds for worst case errors at all levels through equations \ref{eq:levelacc} and \ref{eq:toplevelacc}. Moreover, if the number of subgraphs $g$ is large, the final probability of success if low. As explained in section \ref{sec:results} however, we can select an $\epsilon$ to control both the final accuracy and the probability of success at the expense of logarithmically more transmissions.

\subsection{Is $a = \frac{2}{3} $ optimal? }

We have selected $a=\frac{2}{3}$ in the previous sections to get linear cost at each intermediate hierarchy level. One could ask whether this is the best we can do. Maybe a different choice of $a$ could yield even smaller communication cost. We investigate this question here. As we will see, although $a=\frac{2}{3}$ is not the unique optimal option, it is a well justified choice.

For convenience in the analysis we change the notation a little bit. In Sections~\ref{sec:hgossip} and \ref{sec:analysis} we use $a=\frac{2}{3}$ as a rule for subdividing each cell at one level to its subcells. Here, let us assume that we have subdivision parameters $b_1, b_2,\ldots, b_{k-1}$ with a slightly different meaning. While $a$ is ``local'' and allows transitioning from one level to the next, $b_j$ directly specifies exactly how many cells and nodes in each cell we have at level $j$. Specifically, at level $j$ there will be a total of $n^{1-b_{j-1}}$ cells and each cell will contain $n^{b_{j-1}}$ nodes and $n^{b_{j-1} - b_j}$ leaders communicating over distances of $n^{\frac{b_j}{2}}$ hops. There is a connections between the $b$'s since, if $a$ is constant for all levels then $b_{j} = a^{j}$. We also require that $b_j \geq b_{j-1}$; otherwise, there would be more nodes in a $C_{j+1}$ cell than in a $C_j$ cell which is not consistent with our notion of refining the hierarchical partition. 

Recall from the complexity analysis that the total number of messages $M_j$ at some level $j$ is
\begin{equation}
 M_j = (\#C_j\text{ cells}) \times (\#\text{ leaders per }C_j)^2 \times (\# \text{ hops per message})
\end{equation}
It useful to write the exact expression for the most important cases so:
\begin{itemize}
\item At level $1$ we only have one cell so:
\begin{equation}
M_1 = 1 \cdot (n^{1-b_1})^2 \cdot n^{\frac{b_1}{2}} = n^{2 - \frac{2}{3} b_1} \text{ messages}
\end{equation}

\item At level $1 < j < k$:
\begin{equation}
M_j = n^{1-b_{j-1}} \cdot (n^{b_{j-1}-b_j})^2 \cdot n^{\frac{b_j}{2}} = n^{1 + b_{j-1} - \frac{2}{3} b_j} \text{ messages}
\end{equation}

\item At last level $k$ all messages can be delivered in one hop so:
\begin{equation}
M_k = n^{1-b_{k-1}}  \cdot (n^{b_{k-1}})^2 \cdot 1 = n^{1 + b_{k-1}} \text{ messages}
\end{equation}
\end{itemize}

Notice that even if we select all the $b_j$'s so that $M_j = O(n)$, the $k$-th level will dominate with superlinear complexity since $b_{k-1} > 0$.  Now, if we take a fixed number of levels $k$ we are interested to choose the $b_j$s that minimize the total number of messages $\sum_{j=1}^{k} M_j$ as $n \rightarrow \infty$. This is equivalent to the optimization problem:
%\begin{align}
\begin{eqnarray}
\text{minimize}_{b_1,\ldots, b_{k-1}} & & \max\{2- \frac{2}{3}b_1, 1+b_1 - \frac{2}{3}b_2, \ldots, 1 + b_{k-1}\} \\
\text{subject to   } & & 1 \geq b_1 \geq b_2 \\
& & b_2 \geq b_3 \\
& & \vdots \\
& & b_{k-1} \geq 0.
\end{eqnarray}
%\end{align}
In general the solution to this problem does not yield $a=\frac{2}{3}$ at each level. 

As discussed in section \ref{sec:analysis} to get near linear (e.g. $O\big(n \log \log n \log \frac{kn}{\epsilon} \big)$) complexity we can allow the number of levels $k$ to depend on $n$ so that $k(n) \rightarrow \infty$ as $n \rightarrow \infty$. Notice that even if we use enough levels to have a fixed number of nodes at the finest level, we end up with a linear $O(n)$ number of cells at level $k$ and require constant time to gossip in each. As a result the finest level's complexity can be linear at best. 

If the number of levels is variable, it depends on the values of the subdivision parameters $b_{j}$. If the $b_{j}$'s are large then each cell contains many nodes and we need to use many levels until we reach the finest level. If on the other hand  $b_{j}$'s are small, we create a lot of small cells and few levels. If the number of cells it too large however, we can no longer have $O(n)$ messages at an intermediate level. Consequently, we need to have as few levels as possible. This principle, together with the desire to have as few messages as possible can justify the selection of $a=\frac{2}{3}$ at each level. To see this, let us demand that $M_{j} = O(n) $. For the first level this is true if $2 - \frac{3}{2} b_{1} \leq 1$ which gives $ b_{1} \geq \frac{2}{3}$. Similarly for any other level we need $b_{j} \geq \frac{2}{3} b_{j-1}$. Obviously the smallest possible $b_{j}$s that are still large enough to admit linear complexity at each level are such that $b_{j} = \left( \frac{2}{3}\right)^{j}$. This is the same as using the same $a=\frac{2}{3}$ to subdivide each cell at each level.

% 
% This is can be viewed as a linear program. For example using $3$ levels for simplicity we are interested in solving the following:
% 
% \begin{align}
% \min_{t, b_1, b_2} t & \\
% \text{subject to  } & 2 - \frac{2}{3} b_1 \leq t \\
% & 1+b_1 - \frac{2}{3} b_2 \leq t \\
% & 1 + b_2 \leq t \\
% & b_1 \leq 1 \\
% & b_2 \leq b_1 \\
% & 0 < b_2
% \end{align}

\section{Experimental Evaluation}
\label{sec:evaluation}
In this section we evaluate multiscale gossip in simulation and study its behaviour in practical scenarios. First we investigate the effect of using few versus many levels. Then we show that multiscale gossip performs very well against path averaging \cite{benezit07}, the current state-of-the-art gossip algorithm that requires linear number messages in the size of the network to converge to the average with $\epsilon$ accuracy. Finally, we investigate scenarios where transmissions do not always succeed and messages are either retransmitted or lost.

\subsection{Varying levels of Hierarchy}

\begin{figure} 
\begin{center}
\includegraphics[width=8cm]{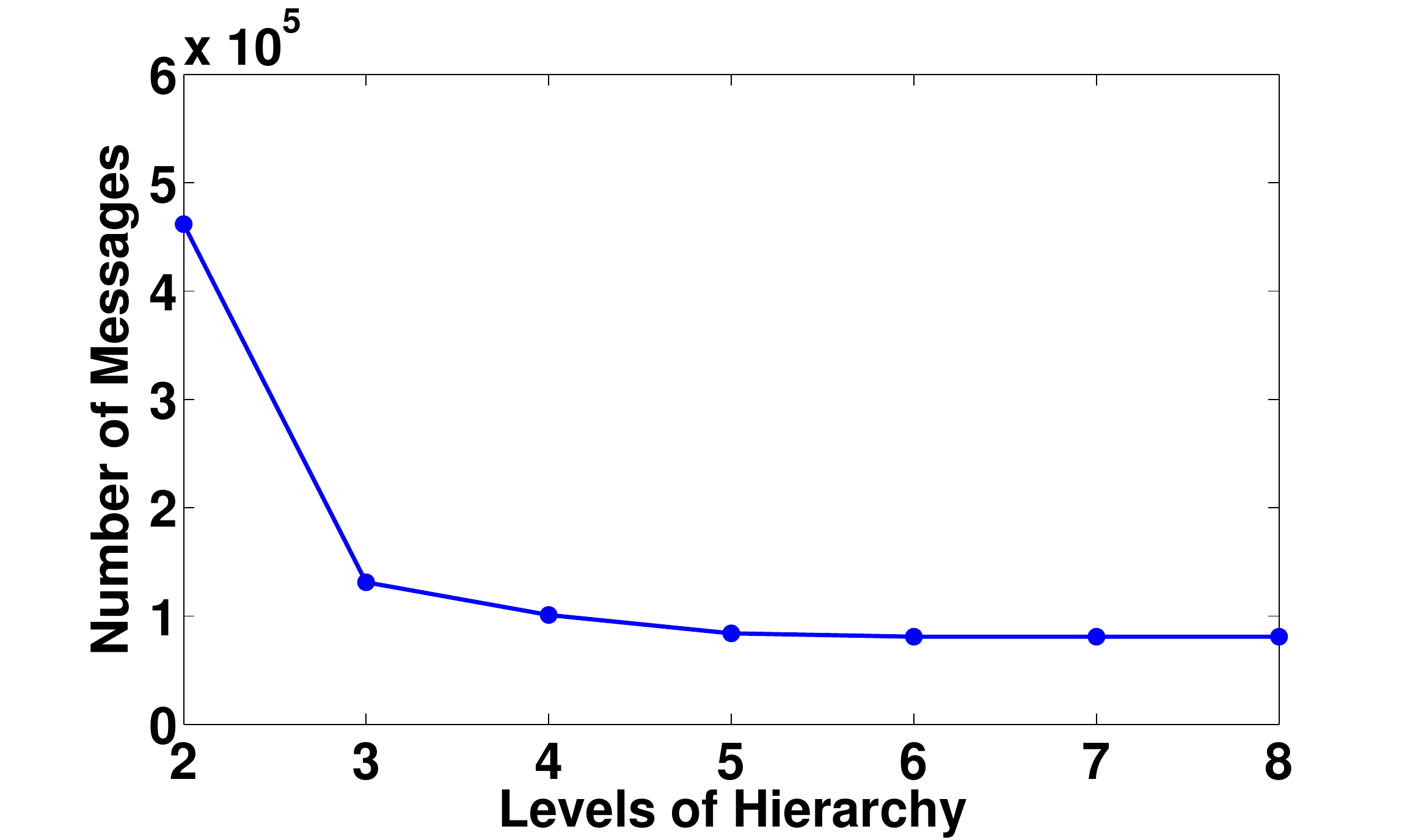}
\caption{\label{fig:manylevels} Increasing the levels of hierarchy yields a diminishing reward. Results are averages over $10$ random geometric graphs with $5000$ nodes and final accuracy $\epsilon = 0.0001$. All graphs are created with radius $r = \sqrt{\frac{3 \log{n}}{n}}$.} 
\end{center}
\end{figure}

In the analysis we concluded that we can select the number of levels $k = \log{\log{n}}$ i.e. we don't need too many levels. This can be verified in practice. Figure \ref{fig:manylevels} investigates the effect of increasing the levels of hierarchy. The figure shows the number of messages until convergence within $0.0001$ error, averaged over ten graphs of  $5000$ nodes. More levels yield a diminishing reward and we do not need more than $4$ or $5$ levels. As discussed in the next subsection this observation led us to try a scheme with only two levels of hierarchy which still produces an efficient algorithm. 

\subsection{Mutliscale Gossip vs Path Averaging}

We compare multiscale gossip against path averaging \cite{benezit07} which is in theory the fastest algorithm for gossiping on random geometric  graphs. Is it worth emphasizing that both algorithms operate under the same two assumptions. First, each nodes needs to know the coordinates of itself and its neighbours on the unit square. Second, each node needs to know the size of the network $n$. In path averaging this is implicit since each message needs to be routed back to the source through the same path. It is thus necessary that nodes have global unique ids which is equivalent to knowing the maximum id and thus the size of the network. In multi scale gossip, the network size is used for each node to determine its role in the logical hierarchy and also decide the number of hierarchy levels.

\begin{figure} 
\begin{center}
\includegraphics[width=8cm]{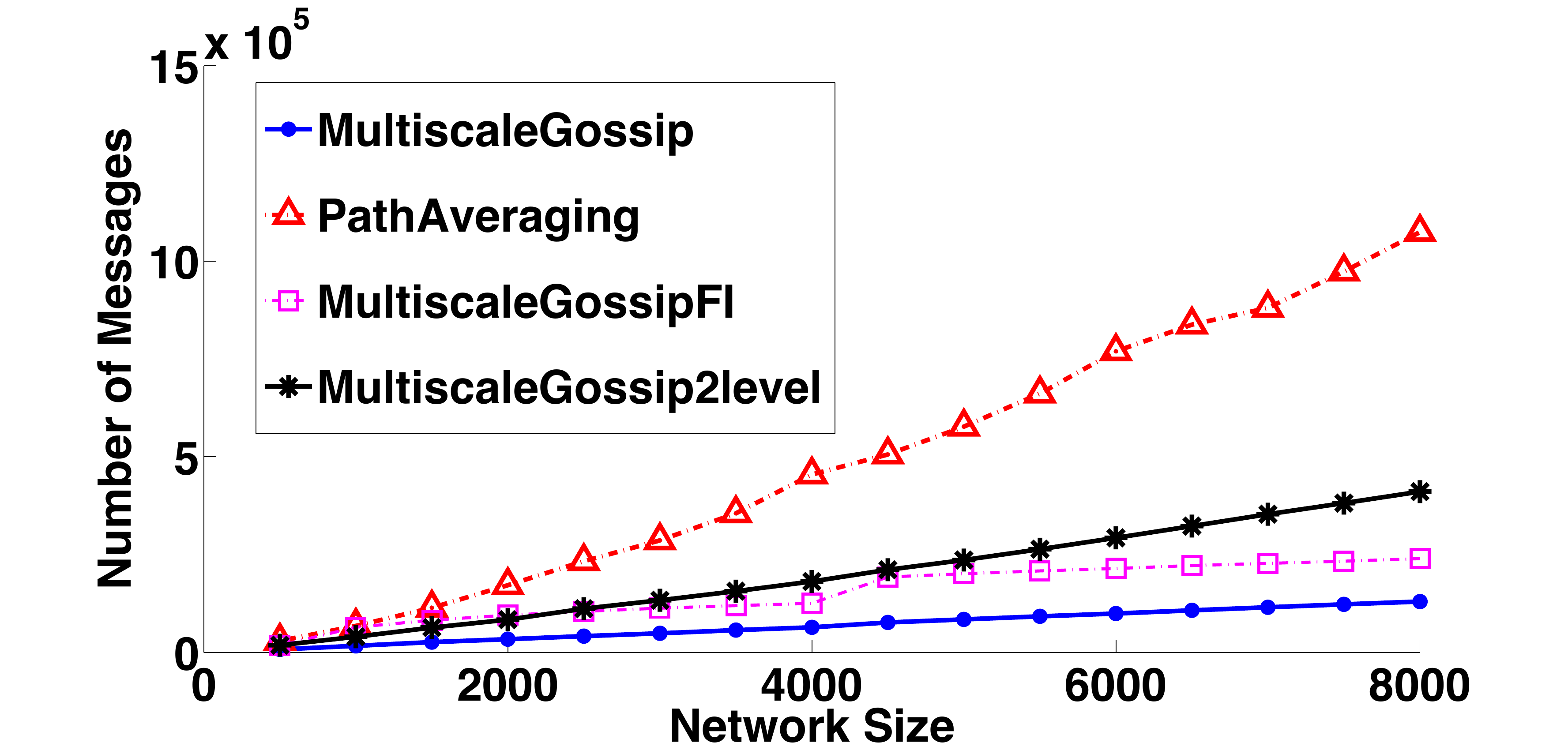}
\caption{\label{fig:hgvspa} Comparison of multiscale gossip to path averaging on random geometric graphs of increasing sizes. MultiscaleGossip used with $5$ levels of hierarchy. MultiscaleGossipFI is the version using a fixed number of iterations for gossiping at a specific level. MultiscaleGossip2level is a version using only two levels of Hierarchy and is explained in Section \ref{sec:consider}. Results are averages over $20$ random geometric graphs with $8000$ nodes and final accuracy $\epsilon = 0.0001$ reached on all runs. All graphs are created with radius $r(n) = \sqrt{\frac{3 \log{n}}{n}}$.} 
\end{center}
\end{figure}

Figure \ref{fig:hgvspa} shows the number of messages needed to converge within $\epsilon=0.0001$ error for graphs of sizes $500$ to $8000$. The bottom curve tagged \textit{MultiscaleGossip} shows the ideal case where computation inside each cell stops automatically when the desired accuracy is reached. The curve labeled \textit{MultiscaleGossipFI} was generated using fixed number of iterations per level based on worst case graph sizes and the curve labeled \textit{MultiscaleGossip2level} was generated using only two levels of hierarchy and an $a=\frac{1}{2}$ instead of $\frac{2}{3}$. Both of these variants are explained below. For path averaging we also simulate the ideal scenario where nodes stop transmitting automatically when achieving the targeted accuracy. As we see all variants of  multiscale gossip use noticeably fewer transmissions than path averaging. One reason why path averaging seems to be slower than in \cite{benezit07} is because we use  a smaller connectivity radius for our graphs ($r = \sqrt{\frac{3 \log{n}}{n}}$ instead of $r = \sqrt{\frac{10 \log{n}}{n}}$). %Moreover, our evaluation of path averaging appears different from the original paper both because our stopping criterion is different and required final accuracy is greater. 

\begin{figure}
\begin{center}
\includegraphics[width=8cm]{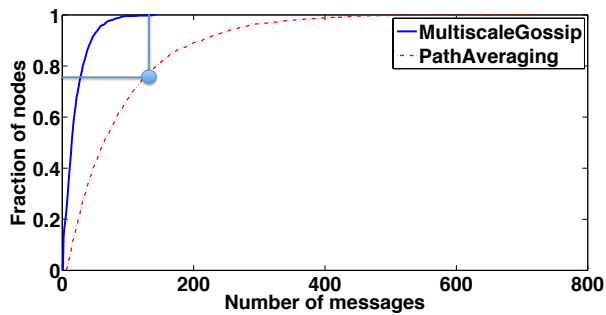}
\caption{\label{fig:cdf} Cumulative density function of the probability that a node sends less of equal to $t$ messages as a function of $t$ on a random geometric graph with $2000$ nodes. The node with maximum number of transmissions for multiscale gossip has less transmissions that about $22\%$ of the nodes in path averaging.} 
\end{center}
\end{figure}

Figure \ref{fig:cdf} depicts the cumulative density functions of transmissions for multiscale gossip and path averaging. Specifically, we plot the fraction of nodes that transmitted $t$ times or less for a random geometric graph with $2000$ nodes. Both path averaging and multiscale gossip were stopped as soon as the desired error level is reached: $\norm{x(t) - \overline{x}_{av}} \leq  0.0001 \norm{x(0)}$. As we see, the node with most transmissions in multiscale gossip still sends fewer messages than about $22\%$ of the nodes in path averaging.  

Multiscale gossip has several advantages over Path Averaging. All the information relies on pairwise messages. In contrast, averaging over paths of length more than two has two main disadvantages. First, if a message is lost, a large number of nodes (potentially $O(\sqrt{\frac{n}{\log{n}}})$) are affected by the information loss. Second, when messages are sent to a remote location over many hops, they increase in size as the message body accumulates the information of all the intermediate nodes. Besides being variable, the message size now depends on the length of the path and ultimately on the network size. Our messages are always of constant size and independent of the hop distance or network size. Moreover, the maximum number of hops any message has to travel is $O(n^{\frac{1}{3}})$ at worst\footnote{At level $1$ the distance in hops between leaders is at worst $O(n^{\frac{a}{2}}) = O(n^{\frac{1}{3}})$ for $a=\frac{2}{3}$.}. This should be compared to distance $O(\sqrt{n})$ which is necessary for path averaging to achieve linear scaling. Finally, multiscale gossip is relatively easy to analyze and implement using standard randomized gossip as a building block for the averaging computations.

\subsubsection*{Fixed Number of Iterations per Level} The ideal scenario for multiscale gossip is if computation inside each cell stops automatically when the desired accuracy is reached. This way no messages are wasted. However in practice for cells at the same level may need to gossip on graphs of different sizes that take different numbers of messages to converge. This creates a need for node synchronization so that all computation in one level is finished before the next level can begin. To alleviate the synchronization issue, we can fix the number of randomized gossip iterations per level so that all computation between different subgraphs at the same level takes practically the same amount of time. However, we need to be careful not to perform fewer iterations than needed for the desired accuracy. Given that nodes are deployed uniformly at random in the unit square, we can make a worst case estimate of how many nodes are expected to be in a cell of a certain area. Since by construction all cells at the same level have equal area, we gossip on all graphs at that level for a fixed number of iterations. Moreover, as seen in the previous section, we can use enough levels of hierarchy to only have $m \leq k \leq M$ nodes at the last level. This can ensure that we will not do less iterations that necessary. In practice, usually at level $k$, we have less nodes than expected so we end up wasting messages running gossip for longer than necessary. 

\subsubsection*{Two-level Gossip} multiscale gossip is a synchronized algorithm where computation in one level begins after the previous level has converged. Synchronization can be complicated or inefficient if we have too many levels. This motivates trying an algorithm with only two levels. In this case, for graphs of size a few thousand nodes, splitting the unit square into $n^{1-a}$ cells with $a=\frac{2}{3}$ is not a good choice as it produces a very small grid of representatives and quite large level-$1$ cells. To achieve better load balancing between the two levels, we use $a=\frac{1}{2}$. This choice has the advantage that the maximum number of hops any message has to travel is $O(n^\frac{1}{4})$. To see this is true, observe that each $C_{(2,\cdot)}$ cell  has area $\frac{1}{n^a} = n^{-\frac{1}{2}} = n^{-\frac{1}{4}} \times n^{-\frac{1}{4}}$. Thus the maximum distance between representatives is $O(n^{-\frac{1}{4}})$. If we divide by the connecting radius $r(n) = \sqrt{\frac{c \log{n}}{n}}$ we get the result. Another interesting finding is that for moderate sized graphs, using cells of area $n^{-\frac{1}{2}}$ produces subgraphs which are very well connected. Since nodes are deployed uniformly at random, an area $n^{-\frac{1}{2}}$ is expected to contain $n^{\frac{1}{2}}$ nodes. A subgraph inside a  $C_{(2,\cdot)}$ cell  is still a random geometric graph with $t=n^{\frac{1}{2}}$ nodes, but for which the radius used to connect nodes is not $\sqrt{\frac{c \log{t}}{t}}$. It is $\sqrt{\frac{c \log{n}}{n}}$. This is equivalent to creating a random geometric graph of $t$ nodes in the unit square but with a scaled up radius of $r_t = \sqrt{\frac{c \log{n}}{t}}$. From \cite{cover_mixing_ran_geom_ercal_07} we know that a random geometric graph of $t$ nodes is rapidly mixing (i.e. linear number of messages for convergence) if the connecting radius is $r_{rapid} = \frac{1}{poly(\log{t})}$. Now, e.g. for $c=3$ and $n \leq 9*10^6$, we get  $r_t \geq \frac{1}{\log{t}} \geq r_{rapid}$ for $t = \sqrt{n} \leq 3000$. Consequently, the $C_{(2,\cdot)}$ cells are rapidly mixing for networks of less than a few millions of nodes. Figure \ref{fig:hgvspa} verifies this analysis. For graphs from $500$ to $8000$ nodes and final error $0.0001$, we see that \textit{MultiscaleGossip2level} performs very close to multiscale gossip with more levels of hierarchy and better than path averaging.

\subsection{Operating under Transmission Failures}

As explained in the previous section, multiscale gossip needs to send messages to shorter distances across the network than Path Averaging. It is important to see what effect this has on the robustness of the algorithms against transmission failures. Two different but general scenarios are considered. In the first scenario, no message is truly lost. There is a non-zero probability that a message over a network edge is not sent successfully, but the nodes communicate via a hand shake mechanism so messages are eventually delivered after a number of attempts. If the probability of successful transmission is $p$, then the cost for a single message transmission over an edge is geometrically distributed : $P[Cost = m] = Geo(p, m)$. The second scenario is more extreme. Each message is delivered with probability $p$ over each edge, otherwise it is lost. This model has  severe consequences. Depending on where in its path the message is lost, a number of nodes will not update their values properly so besides the overall delay in convergence, part of the signal energy is lost and the final estimate of the average is no longer guaranteed to be close to the true average. 

\subsubsection{Hand Shake Model}

\begin{figure} [t]
\begin{center}
\includegraphics[width=8cm]{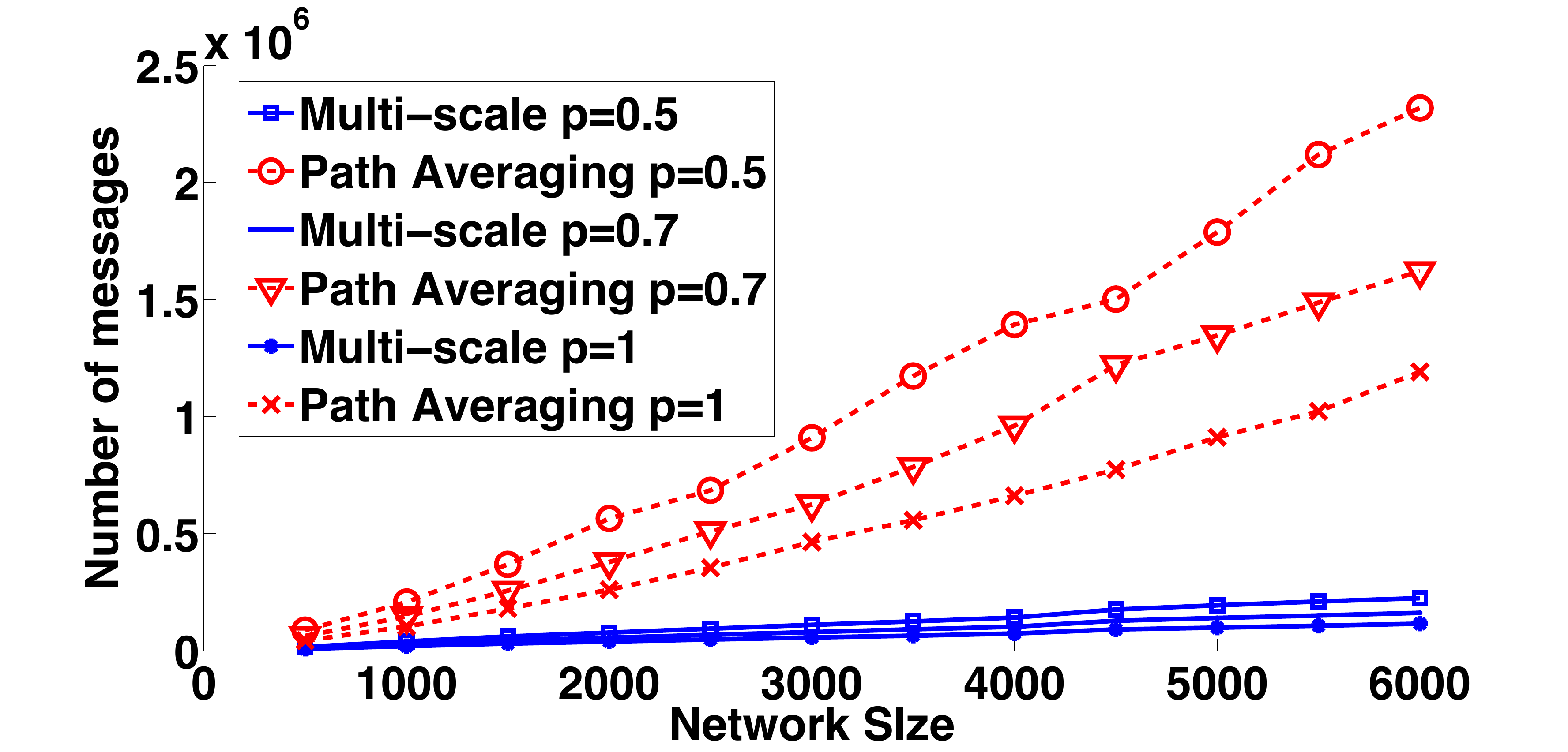}
\caption{\label{fig:handshake} Multiscale gossip and Path averaging on graphs of varius sizes using hand shakes to overcome transmission failures. The probability of successful transmission $p$ varies from $0.5$ to $1$. All results are averaged over 25 graphs of each size. Final accuracy is $\epsilon = 0.0001$ and the random geometric graphs use $r(n) = \sqrt{\frac{3 \log{n}}{n}}$.} 
\end{center}
\end{figure}

In this scenario we don not have to worry about convergence. All messages are eventually delivered and it is just a matter of time. Figure \ref{fig:handshake} shows the results of multiscale gossip against path averaging on networks of different sizes. The probability of successful transmission also varies from $p=0.5$ to $p=1$. As we see multiscale gossip is significantly less affected by such failures. This example clearly illustrates the importance of not having to send messages in long distances of the network. Since each individual link introduces some delay, the fact that messages in multiscale gossip usually don't need to travel far and need to go $O(n^{\frac{1}{3}})$ hops at most, allow the algorithm to converge using much fewer messages than path averaging.

\subsubsection{Message Loss Model}

In this scenario, a message is delivered with probability $p$ or lost forever. This severely impedes the algorithms from converging fast. Moreover, information is lost permanently distorting the final result and making it impossible to meet the desired final accuracy. The amount of distortion depends on where along a multi-hop path the failure occurs. For example in path averaging, if the message is lost at the first transmission on its way back, all the nodes along the path except the last will have distorted information. Similar situations occur in multiscale gossip between leader communications where messages need to travel across multiple hops. Since in this scenario we have no guarantee that the desired final accuracy criterion will be met, it is hard to draw conclusions whether multiscale gossip is better than path averaging. Our observations showed that both algorithms can only reach an accuracy in the order of $0.01$ when targeting at $\epsilon = 0.0001$. Specifically, multiscale gossip would only reach up to $0.06$ accuracy while path averaging could not improve beyond $0.02$.  At the same time the total number of messages still increases linearly for multiscale gossip while it seem to blow up exponentially for path averaging.

%  As we see in Figure \ref{fig:fail} multi-scale gossip still uses orders of magnitude fewer messages. This is explained by the fact that the maximum distance any message has to travel is $O(n^{\frac{1}{3}})$ for multi-scale gossip which is shorter than $O(\sqrt{n})$ required by path averaging. However, as the computation happens in stages, any distrortion suffered due to lost messages affects all the higher levels and the overall distortion is more severe. As we see in figure \ref{fig:failerr} for multiscale gossip we can no longer reach the $\epsilon = 0.0001$ accuracy. Both algorithms yield much coarsed approximations to the average, but the distortion for Path Averaging is less. 

% \begin{figure}
% \begin{center}
% \includegraphics[width=8cm]{./figures/Fail.pdf}
% \caption{\label{fig:fail} Number of messages for multi-scale gossip and Path averaging on graphs of varius sizes when messages are lost forever. The probability of successful transmission $p$ varies from $0.5$ to $1$. All results are averaged over 25 graphs of each size.} 
% \end{center}
% \end{figure}
% 
% 
% \begin{figure}
% \begin{center}
% \includegraphics[width=8cm]{./figures/failerr.pdf}
% \caption{\label{fig:failerr} Final error of multi-scale gossip and Path averaging on graphs of varius sizes when messages are lost forever. The probability of successful transmission $p$ varies from $0.5$ to $1$. All results are averaged over 25 graphs of each size.} 
% \end{center}
% \end{figure}

\section{Practical Considerations}
\label{sec:consider}
There is a number of practical considerations that we would like to bring to the reader's attention. We list them in the form of questions below:

\textbf{How can we detect convergence in a subgraph or cluster? Do the nodes need to be synchronized?} At each hierarchy level, representatives know how big the grid that they are gossiping over is (function of $n$ and $k$ only). Moreover, all grids at the same level are of the same size and we have tight bounds on the number of messages needed to obtain $\epsilon$ accuracy on grids w.h.p.  We can thus gossip on all grids for a fixed number or rounds and synchronization is implicit.  At level $k$ however, in general we need to gossip on random geometric subgraphs which are not of exactly the same size. As $n$ gets large though, random geometric graphs tend to become regular and uniformly spaced on the unit square. Therefore, the subgraphs contained in cells at level $k$ all have sizes very close to the expected value of $n^{(\frac{2}{3})^{k}}$.  Thus, we run gossip for a fixed number of rounds using the theoretical bound for graphs of the size $n^{(\frac{2}{3})^{k}}$. As discussed in Section \ref{sec:hgossip}, fixing the number of iterations leads to redundant transmissions, however the algorithm is still very efficient. 

\textbf{What happens with disconnected subgraphs or grids due to empty grid cells?} This is possible since the division of the unit square into grid cells does not mean that each cell is guaranteed to contain any nodes of the initial graph. Representatives use multi-hop communication and connected grids can always be constructed as long as the initial random geometric graph is connected. At level $k$ the subgraphs of the initial graph contained in each cell could still be disconnected if edges that go outside the cell are not allowed. However, as explained in Section \ref{sec:analysis} we can use enough hierarchy levels so that each $C_{(k,\cdot)}$ cell is a complete graph and the probability of getting disconnected $C_{(k,\cdot)}$ cells tends to zero.

\textbf{How can we select representatives in a natural way?} The easiest solution is to pick the point $p_{c}$ that is geographically at the center of each cell. Again, knowledge of $n,k$ uniquely identifies the position of each cell and also $p_{c}$. By sending all messages to $p_{c}$,  geographic routing will deliver them to the unique node that is closest to that location w.h.p. To change representatives, we can deterministically pick a location $p_{c} + u$ which will cause a new node to be the closest to that location. A more sophisticated solution would be to employ a randomized auction mechanism. Each node in a cell generates a random number and the largest number is the representative. Once a new message enters a cell, the nodes knowing their neighbours' values, route the message to the cell representative. Notice that determining cell leaders this way  does not incur more than linear cost.

\textbf{Are representatives bottlenecks and single points of failure?} This is not an issue. There might be a small imbalance in the amount of work done by each node, but it can be alleviated by selecting different representatives at each hierarchy level. Moreover, for increased robustness, at a linear cost we can disseminate the representative's values to all the nodes in its cell. This way if a representative dies, another node in the cell can take its place. The new representative will have a value very similar (within $\epsilon$) to that of the initial representative at the beginning of the computation at the current level. Thus node failure is expected to only cause small delay in convergence at that level. We should emphasize however that the effect of node failures has received little attention so far and still asks for a more systematic investigation.

\begin{figure}
\begin{center}
\includegraphics[width=8cm]{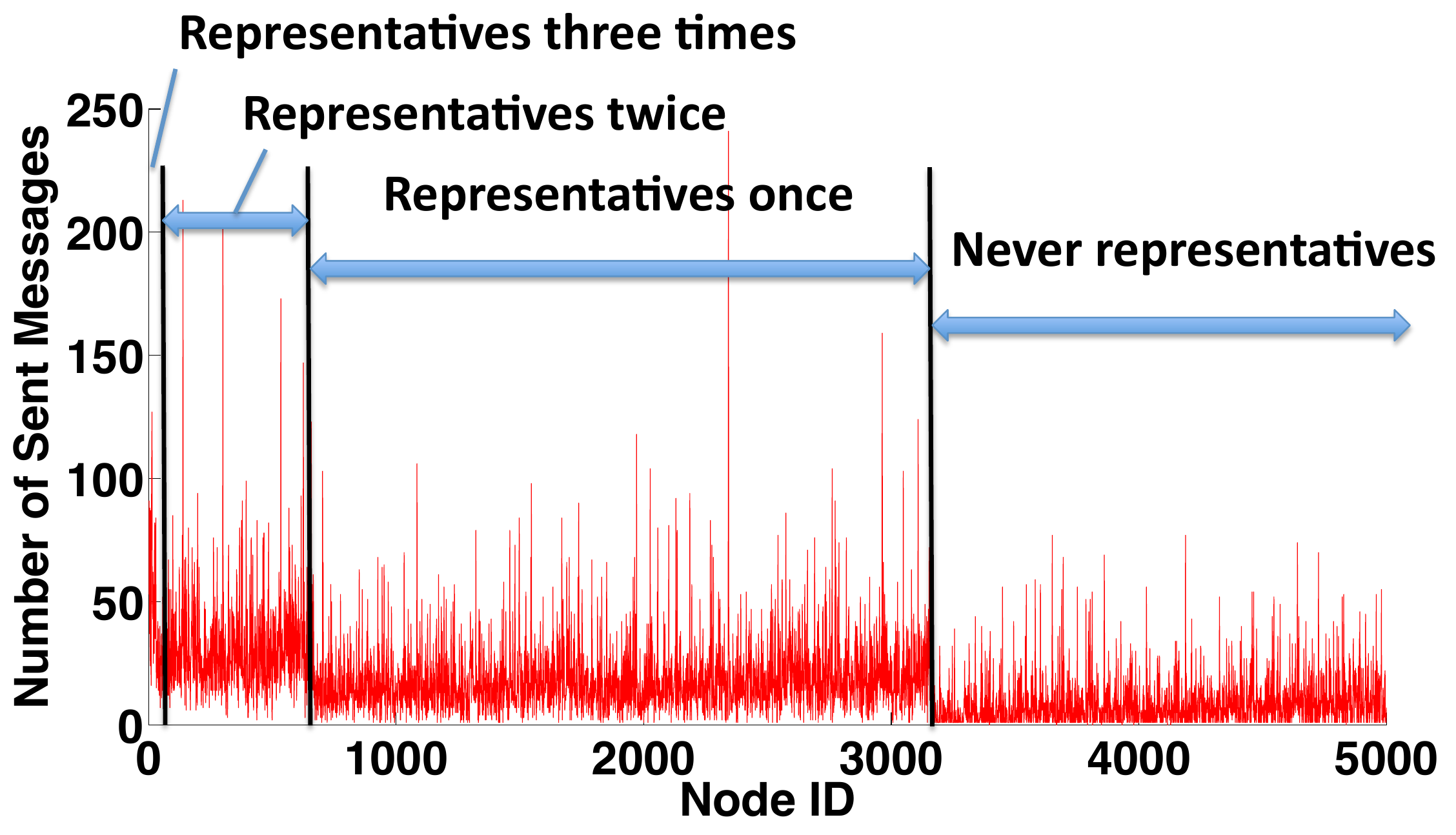}
\caption{\label{fig:nodeutil} Node utilization on a random geometric graphs with $5000$ nodes and final desired accuracy $\epsilon=0.0001$. For each node the total number of sent messages is shown. Nodes $1$ to $30$ have been representatives three times. Nodes $31$ to $662$ have been representatives two times. Nodes $663$ to $3157$ have been representatives once. The rest of the nodes never had a representative role. All nodes may have participated as intermediates in multi-hop communication.} 
\end{center}
\end{figure}

\textbf{How much extra energy do the representatives need to spend? } This question is difficult to answer analytically. We use simulation to get a feel for it. Figure \ref{fig:nodeutil} shows the number of messages sent by each of the $5000$ nodes in a random geometric graph. For this case we used five levels of hierarchy. We expect that some nodes will transmit more messages since, as we move down the hierarchy, cells get smaller and there are fewer nodes from which to draw representatives. In this example, by randomly selecting representatives at each level, no node was a representative more than $3$ times. We show the number of transmissions for nodes of each type in table \ref{tab:nodeutil}, including messages relayed by intermediate nodes using geographic routing. As we see, most of the nodes use very few messages. Moreover, the average degree of this particular example is $26$, and thus, on average each node sends fewer messages than it has neighbors.

\begin{table} 
\label{tab:nodeutil}
\caption{Mean and standard deviation of number of transmissions for different types of nodes running multiscale gossip on a random geometric graph with $5000$ nodes with $5$ levels of hierarchy (See also figure \ref{fig:nodeutil}). }
\begin{center}
\begin{tabular}{|c|c|c|}
\hline
\textbf{Node type} & \textbf{Mean \#msg} & \textbf{Std}\\
\hline
Three times representatives & $58.63$ & $27.05$\\
\hline
Two times representatives & $31.23$ & $21.11$ \\
\hline
One time representatives & $18.77$ & $15.21$ \\
\hline
Never representatives & $8,65$ & $10.23$ \\
\hline
All nodes & $16.85$ & $16.67$ \\
\hline
\end{tabular}
\end{center}
\end{table}

\begin{figure}
\begin{center}
\includegraphics[width=8cm]{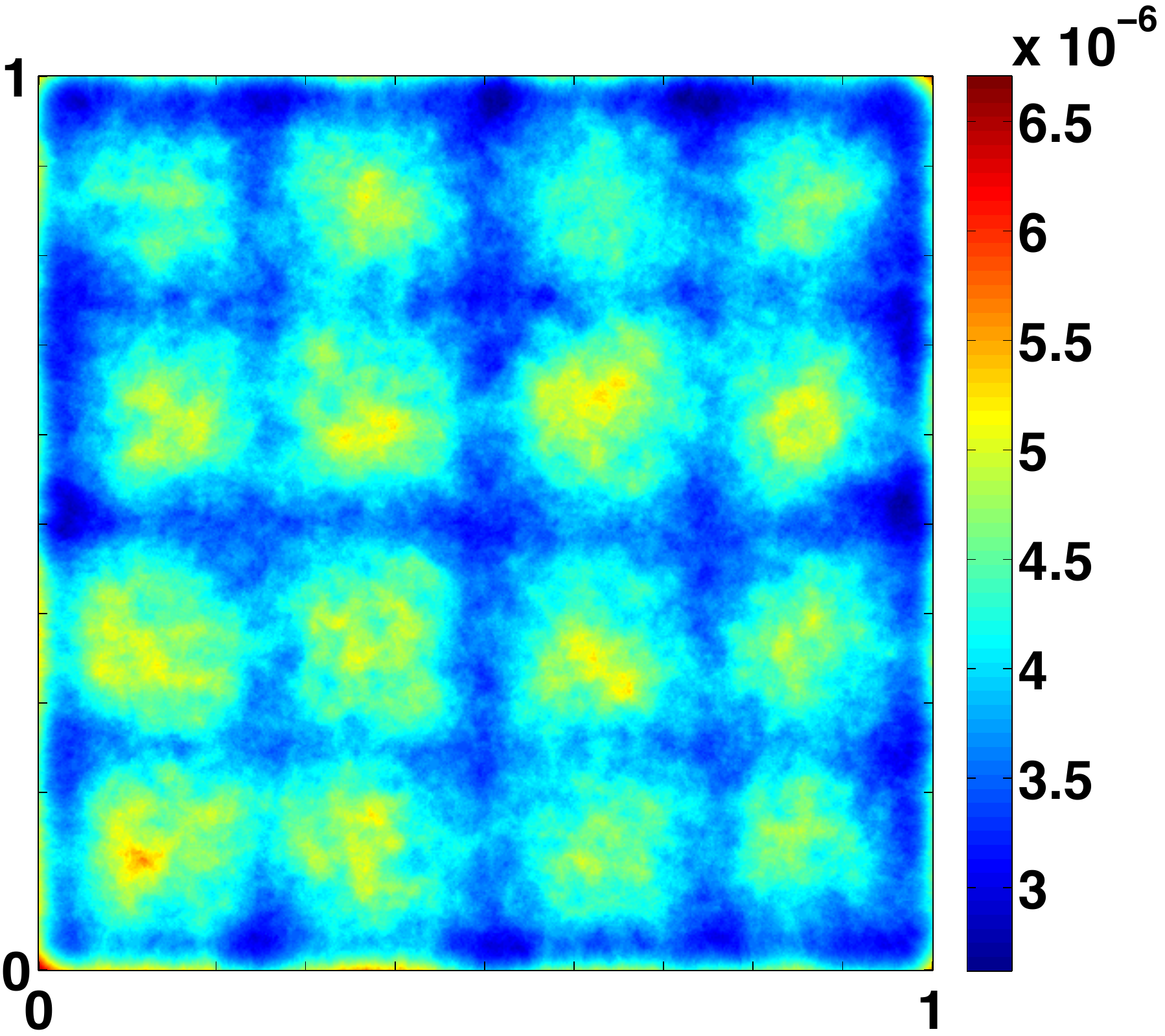}
\includegraphics[width=8cm]{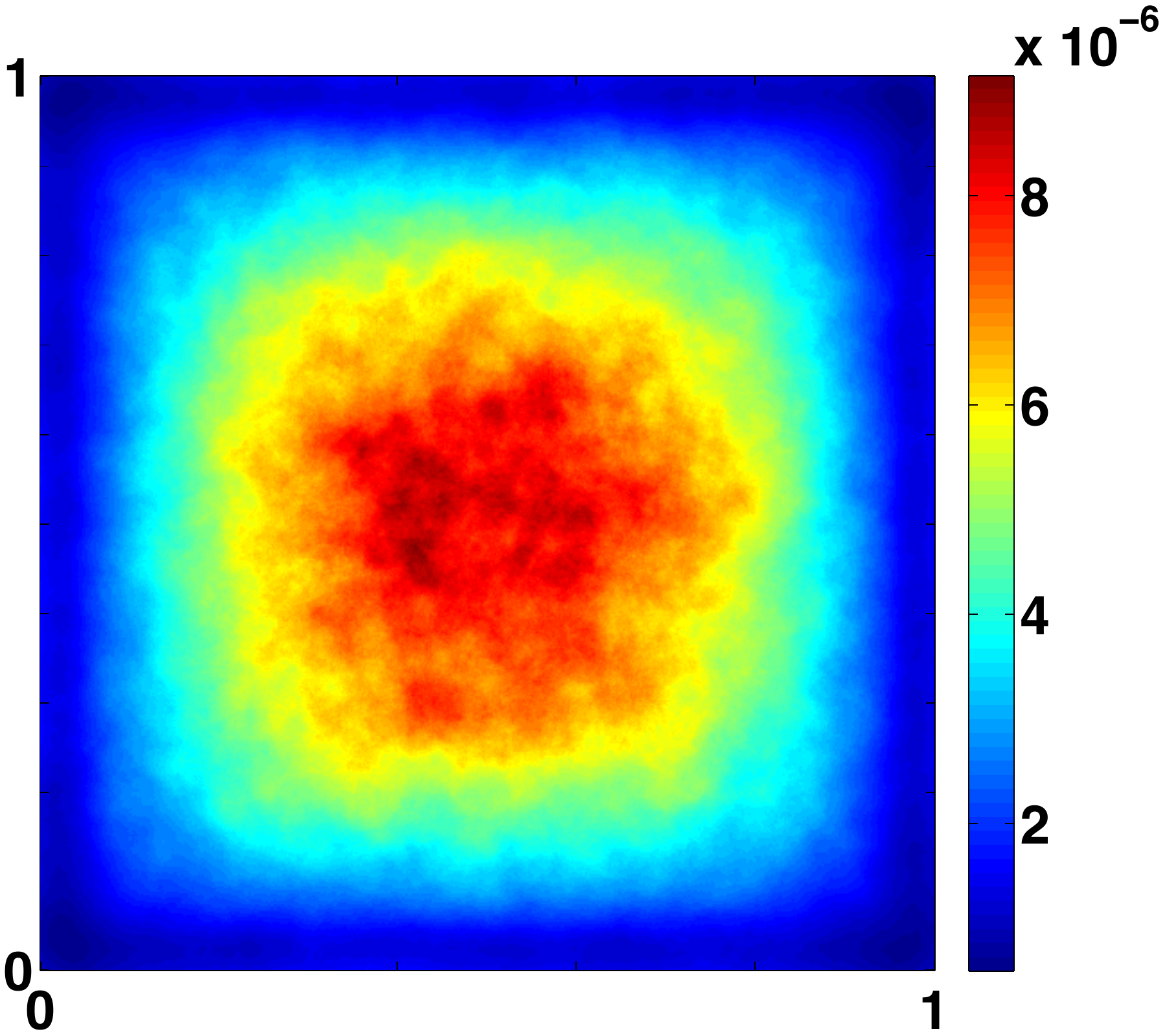}
\caption{\label{fig:nodeload} Comparison of communication load between multiscale gossip (left) and path averaging (right). Color intensity is proportional to amount of transmissions through that part of the unit square. More blue (lower intensity) means little traffic while more red (higher intensity) means heavy traffic. Multiscale gossip distributes traffic relatively uniformly on all nodes while in path averaging the nodes at the center handle much more traffic than nodes around the perimeter. Results are averages over $100$ graphs of $1000$ nodes each.} 
\end{center}
\end{figure}

Figure \ref{fig:nodeload} illustrates the typical fraction of transmissions per node as a function of location in the unit square normalized as a probability distribution. Each pixel is assigned an intensity proportional to the number of messages that nodes located in that region will typically transmit. Thus this figure reveals which nodes suffer from the heaviest traffic. The figure is the result of averaging over $100$ realizations of gossip for each of $100$ different $1000$-node random geometric graphs. As the figure shows, multiscale gossip tends to distribute traffic almost uniformly in all the nodes. The observed color pattern is consistent with the hierarchical nature of the algorithm and although nodes that become representatives send more messages, no node is heavily used. On the other hand, for path averaging we observe that the nodes at the center (red region) send many more messages than nodes around the perimeter. This should be expected since geographic routing (which path averaging relies on) is greedy when trying to deliver messages to remote locations across the network.

\section{Discussion and Future Work}
\label{sec:future}
We have presented a new algorithm for distributed averaging exploiting hierarchical computation. Multiscale gossip separates the computation in linear phases and achieves very close to linear complexity overall. The key to achieving nearly-linear scaling lies in the way the recursive network partition is constructed. In particular, we argue that refining the network so that subnetworks at scale $j$ contain $O(n^{(2/3)^{j}})$ nodes provides optimal message-scaling laws with minimal number of levels in the hierarchy; although other hierarchical partitioning schemes could be constructed to achieve nearly-linear message complexity, they would require deeper hierarchies and, consequently, additional overhead for management. Our analysis focused on network topologies modeled as random geometric graphs, but it should be clear that the results translate directly to grids ($2$-dimensional lattices) embedded in the unit square.

Another feature of the proposed scheme is that the maximum distance any message has to travel is $O(n^{\frac{1}{3}})$ hops, which is shorter than $O(\sqrt{n/\log(n)})$ hops needed by path averaging~\cite{benezit07}, where each iteration involves averaging along a path of nodes potentially spanning the diameter of the network. Requiring transmissions over shorter distances is advantageous when transmitting over unreliable links that use acknowledgements and retransmission to ensure reliable communication at the link-layer, as is common practice in many existing systems. Intuitively, shorter paths translates directly to fewer retransmissions, and we illustrate this via simulation.

There is a number of interesting future directions that we see. In our present algorithm, gossip at higher levels happens on overlay grids which are known to require a number of messages which scales quadratically in the size of the grid. Since these grids already use multi-hop communication, it may be possible to further increase performance by devising other overlay graphs between representatives with better convergence properties, i.e. expander graphs \cite{Margulis_expanders}. Moreover, the subdivision of the unit square into a grid cell is not necessarily natural with respect to the topology of the graph, and one could use other methods for constructing hierarchical partitions which are tuned to the network topology.  Our preliminary results with using hierarchical spectral clustering appear promising in simulation. It is, however, not clear how to carry out spectral clustering in a decentralized manner way and in linear number of messages. Another possibility is to combine the multiscale approach with the use of more memory at each node to get faster mixing rates. Notice however that how to use memory to provably accelerate asynchronous gossip is still an open question. Current results only consider synchronous algorithms~\cite{Oreshkin10}. Finally, an important advantage of gossip algorithms in general is their robustness. However, the general question of modeling and reacting to node failures has not been formally investigated in the literature. It would be very interesting to introduce failures and see the effect on performance for different gossip algorithms.

%\begin{figure}[!t]
%\centering
%\includegraphics[width=2.5in]{myfigure}
% where an .eps filename suffix will be assumed under latex, 
% and a .pdf suffix will be assumed for pdflatex; or what has been declared
% via \DeclareGraphicsExtensions.
%\caption{Simulation Results}
%\label{fig_sim}
%\end{figure}

% Note that IEEE typically puts floats only at the top, even when this
% results in a large percentage of a column being occupied by floats.

% An example of a double column floating figure using two subfigures.
% (The subfig.sty package must be loaded for this to work.)
% The subfigure \label commands are set within each subfloat command, the
% \label for the overall figure must come after \caption.
% \hfil must be used as a separator to get equal spacing.
% The subfigure.sty package works much the same way, except \subfigure is
% used instead of \subfloat.
%
%\begin{figure*}[!t]
%\centerline{\subfloat[Case I]\includegraphics[width=2.5in]{subfigcase1}%
%\hfil
%\subfloat[Case II]{\includegraphics[width=2.5in]{subfigcase2}%
%\label{fig_second_case}}}
%\caption{Simulation results}
%\label{fig_sim}
%\end{figure*}
%
% Note that often IEEE papers with subfigures do not employ subfigure
% captions (using the optional argument to \subfloat), but instead will
% reference/describe all of them (a), (b), etc., within the main caption.

\appendices
\section{Complementary derivations for multi-scale gossip error bound}

Looking at inequality \eqref{eq:mainbound}, we have three terms that we need to bound.

\begin{itemize}
\item $T_{1} = \frac{\epsilon^2 \norm{\{ x_{1:L_1} \}}^2}{\norm{x}^2}$ : We can bound the numerator if we consider the following; At some level $j$ we have $x_{s_{1} \dots s_{j}} = x_{s_{1} \dots s_{j} c_{j+1}}^{*} \leq \max_{ 1 \leq i_{j+1} \leq L_{j+1}} (x_{s_{1} \dots s_{j} i_{j+1}})    $ for some $ 1 \leq c_{j+1} \leq L_{j+1}$. We can keep bounding all the way down to level $k$ in the exact same fashion since $ x_{s_{1} \dots s_{j} i_{j+1}}  \leq x_{s_{1} \dots s_{j} i_{j+1} c_{j+2} }^{*}  \leq \max_{ 1 \leq i_{j+2} \leq L_{j+2}} (x_{s_{1} \dots s_{j} i_{j+1} i_{j+2}})$ for $ 1 \leq c_{j+2} \leq L_{j+2}$ and so on. If we do this for all terms appearing in the vector $\{ x_{1:L_1} \}$, we get a numerator as a norm of elements from the initial $x$-vector. Since we need to divide expression by $\norm{x}$, the ratio is less than one simply since the denominator has more terms. This means  $T_{1} \leq \epsilon^{2}$.

\item $T_{2} = \frac{L_1 (m^{L_1} - x_{av})^2}{\norm{x}^2}$ : Using definition \eqref{eq:mean} and the fact that an average such as $x_{av}$ can be written as the average of averages, 
\begin{equation}
%\frac{1 }{\norm{x}^2} \frac{   \sum_{s_{1}=1}^{L_{1}}  x_{s_{1}}   } {L_{1}}
\frac{m^{L_1} - x_{av}}{\norm{x}} = \frac{1 }{\norm{x}} \left( \frac{\sum_{s_{1}=1}^{L_{1}}  x_{s_{1}}}{L_{1}} - \frac{ \sum_{s_{1}=1}^{L_{1}} \dots \sum_{s_{k}=1}^{L_{k}} x_{s_{1} \dots s_{k}} }{L_{1} L_{2} \cdots L_{k}} \right).
\end{equation}
Pulling $\frac{1}{L_{1}}$  and the summation over $L_{1}$ out, adding and subtracting the $L_{2}$ mean, taking the absolute value and using triangular inequality
\begin{eqnarray}
\frac{m^{L_1} - x_{av}}{\norm{x}} & \leq &\frac{1 }{\norm{x}}  \frac{1}{L_{1}}  \sum_{s_{1}=1}^{L_{1}}  \left( \abs{ x_{s_{1}} - m_{s_{1}}^{L_{2}} + m_{s_{1}}^{L_{2}} - \frac{ \sum_{s_{2}=1}^{L_{2}} \dots \sum_{s_{k}=1}^{L_{k}} x_{s_{1} \dots s_{k}} }{L_{2} L_{3} \cdots L_{k}}    } \right)  \\
& \leq & \frac{1 }{\norm{x}}  \frac{1}{L_{1}}  \sum_{s_{1}=1}^{L_{1}}  \left( \abs{ x_{s_{1}} - m_{s_{1}}^{L_{2}}} + \abs{m_{s_{1}}^{L_{2}} - \frac{ \sum_{s_{2}=1}^{L_{2}} \dots \sum_{s_{k}=1}^{L_{k}} x_{s_{1} \dots s_{k}} }{L_{2} L_{3} \cdots L_{k}}    } \right).
\end{eqnarray}
Using bound \eqref{eq:levelacc}
\begin{equation}
\frac{m^{L_1} - x_{av}}{\norm{x}} \leq \frac{1 }{\norm{x}}  \frac{1}{L_{1}}  \sum_{s_{1}=1}^{L_{1}}  \left( \epsilon \norm{\{ x_{s_{1} 1:L_{2}} \}} + \abs{m_{s_{1}}^{L_{2}} - \frac{ \sum_{s_{2}=1}^{L_{2}} \dots \sum_{s_{k}=1}^{L_{k}} x_{s_{1} \dots s_{k}} }{L_{2} L_{3} \cdots L_{k}}    } \right).
\end{equation}
Using definition \eqref{eq:mean}
\begin{equation}
\frac{m^{L_1} - x_{av}}{\norm{x}} \leq \frac{1 }{\norm{x}}  \frac{1}{L_{1}}  \sum_{s_{1}=1}^{L_{1}} \left( \epsilon \norm{\{ x_{s_{1} 1:L_{2}} \}} + \abs{  \frac{ \sum_{s_{2}=1}^{L_{2}} x_{s_{1} s_{2}}  }{L_{2}} - \frac{ \sum_{s_{2}=1}^{L_{2}} \dots \sum_{s_{k}=1}^{L_{k}} x_{s_{1} \dots s_{k}} }{L_{2} L_{3} \cdots L_{k}}    } \right).
\end{equation}
Pulling the summation and the term $\frac{1}{L_{2}}$ outside
\begin{equation}
\frac{m^{L_1} - x_{av}}{\norm{x}} \leq \frac{1 }{\norm{x}}  \frac{1}{L_{1}}  \sum_{s_{1}=1}^{L_{1}}  \left( \epsilon \norm{\{ x_{s_{1} 1:L_{2}} \}} +  \frac{1}{L_{2}}  \sum_{s_{2}=1}^{L_{2}}  \abs{  x_{s_{1} s_{2}}  - \frac{ \sum_{s_{3}=1}^{L_{3}} \dots \sum_{s_{k}=1}^{L_{k}} x_{s_{1} \dots s_{k}} }{L_{3} L_{4} \cdots L_{k}}    } \right).
\end{equation}
We repeatedly pull out terms and add and subtract means to reach 
\begin{eqnarray} \label{eq:T2}
\frac{m^{L_1} - x_{av}}{\norm{x}} &\leq& \frac{1 }{\norm{x}}  \frac{1}{L_{1}}   \sum_{s_{1}=1}^{L_{1}}  \left( \epsilon \norm{\{ x_{s_{1} 1:L_{2}} \}} +  \frac{1}{L_{2}}  \sum_{s_{2}=1}^{L_{2}} ( \epsilon \norm{\{  x_{s_{1} s_{2} 1:L_{3}} \}} + \dots \right. \\
& & \dots \left. + \frac{1}{L_{k-1}} \sum_{s_{k-1}=1}^{L_{k-1}} ( \epsilon \norm{\{ x_{s_{1} \dots s_{k-1} 1:_{L_{k}}}  \}}  ) + \abs{m_{s_{1}\dots s_{k-1}}^{L_{k}} - \frac{ \sum_{s_{k}=1}^{L_{k}} x_{s_{1} \dots s_{k}} }{L_{k}}  } ) \dots \right) \\
& = &\frac{1 }{\norm{x}}  \frac{1}{L_{1}}   \sum_{s_{1}=1}^{L_{1}}  \left( \epsilon \norm{\{ x_{s_{1} 1:L_{2}} \}} +  \frac{1}{L_{2}}  \sum_{s_{2}=1}^{L_{2}} ( \epsilon \norm{\{  x_{s_{1} s_{2} 1:L_{3}} \}} + \dots  \right. \\
& &\left. \dots + \frac{1}{L_{k-1}} \sum_{s_{k-1}=1}^{L_{k-1}} ( \epsilon \norm{\{ x_{s_{1} \dots s_{k-1} 1:_{L_{k}}}  \}}  ) + 0 ) \dots \right). 
\end{eqnarray}
The comment from bounding $T_{1}$ helps us here as well. Each term in the numerator can be bounded by terms from the initial vector and dividing by $\norm{x}$ we get
\begin{equation}
\begin{split}
\frac{m^{L_1} - x_{av}}{\norm{x}}  \leq \frac{1}{L_{1}} \sum_{s_{1}=1}^{L_{1}} (\epsilon + \frac{1}{L_{2}} \sum_{s_{2}=1}^{L_{2}} (\epsilon + \dots + \frac{1}{L_{k-1}} \epsilon ) \dots ) = (k-1) \epsilon.
\end{split}
\end{equation} 
This is true since each summation cancels out with the corresponding denominator leaving only an $\epsilon$ term and we have $k-1$ such terms. Now obviously $T_{2} \leq L_{1} (k-1)^{2} \epsilon^{2}$

\end{itemize}

\section*{Acknowledgment}

This work is supported by grants from the \emph{Fonds qu\'eb\'ecois de la recherche sur la nature et les technologies}, and the Mathematics of Information Technology and Complex Systems (MITACS) Canadian network of centers of excellence.

The authors would like to thank Marius Sucan for providing the multi-scale grid figure used in Section~\ref{sec:hgossip}.


\bibliographystyle{./IEEEtranBST/IEEEtran}
\bibliography{gossip}

%\begin{biography}[{\includegraphics[width=1in,height=1.25in,clip,keepaspectratio]{mshell}}]{Michael Shell}

%\end{IEEEbiography}
%% insert where needed to balance the two columns on the last page with
%%\newpage


%\vfill


\end{document}